\documentclass[12pt,preprint]{aastex}
\usepackage{amsmath}
\usepackage{amssymb}

\newcommand{\etal}{et~al.~}                                        
\newcommand{\sersic}{S\'{e}rsic\ }

\newcommand{\kms}{\ifmmode\,{\rm km}\,{\rm s}^{-1}\else km$\,$s$^{-1}$\fi} 
\newcommand{\be}{\begin{equation}}
\newcommand{\ee}{\end{equation}}
\newcommand{\bea}{\begin{eqnarray}}
\newcommand{\eea}{\end{eqnarray}}

\newcommand{\pmd}{P(M|D,I)}
\newcommand{\pdm}{P(D|M)}
\newcommand{\prm}{P(M)}
 
\def \spose#1{\hbox to 0pt{#1\hss}}                                   
\def \ltsim{\mathrel{\spose{\lower 3pt\hbox{$\sim$}}                  
     \raise 2.0pt\hbox{$<$}}}                                                 
\def \gtsim{\mathrel{\spose{\lower 3pt\hbox{$\sim$}}                   
     \raise 2.0pt\hbox{$>$}}}

\def\Eq#1{Eq.~(\ref{eq:#1})}
\def\se#1{\S\ref{sec:#1}}
 
\def\Fig#1{Figure~\ref{fig:#1}}

\def\ifm#1{\relax\ifmmode#1\else$\mathsurround=0pt #1$\fi}  
\def\kms{\ifmmode\,{\rm km}\,{\rm s}^{-1}\else km$\,$s$^{-1}$\fi}

\def\msol{M_{\odot}}  
\def\lsol{L_{\odot}}

\def\ltsima{$\; \buildrel < \over \sim \;$}                    
\def\lsim{\lower.5ex\hbox{\ltsima}}  
\def\gtsima{$\; \buildrel > \over \sim \;$}                    
\def\gsim{\lower.5ex\hbox{\gtsima}}

\def\C28{\rm C_{28}} 

\def\pmb#1{\setbox0=\hbox{#1}%
\kern-.025em\copy0\kern-\wd0 \kern.05em\copy0\kern-\wd0 
\kern-.025em\raise.0433em\box0}

\def \littlemm{\ifmmode{\scriptscriptstyle m } 
     \else{\hbox{$\scriptscriptstyle m $ }}\fi}  
\def \topemm{\raise .9ex \hbox{\littlemm}}  
\def \magpoint{\hbox to 2pt{}\rlap{\hskip -.5ex 
     \topemm}.\hbox to 2pt{}} 
\def \deg {$^\circ$} 
\def \magarc {mag arcsec$^{-2}$}

\usepackage{apjfonts}
\usepackage{lscape}

\slugcomment{Accepted for publication at ApJ} 

\shorttitle{The Luminosity Profile of M31} 
\shortauthors{Courteau \etal 2011}

\begin{document}


\title{The Luminosity Profile and Structural Parameters of the Andromeda Galaxy}

\author{St\'ephane Courteau and Lawrence M. Widrow}    
\affil{Department of Physics, Engineering Physics \& Astronomy, Queen's
  University, Kingston, Ontario, Canada}

\author{Michael McDonald}
\affil{Kavli Institute for Astrophysics and Space Research, MIT, Cambridge, MA}

\author{Puragra Guhathakurta} 
\affil{UCO/Lick Observatory and Department of Astronomy and Astrophysics, 
       University of California, Santa Cruz, CA}

\author{Karoline M. Gilbert$^\dagger$} 
\affil{Department of Astronomy, University of Washington, Seattle, WA
       and $\dagger$ Hubble Fellow}

\author{Yucong Zhu} 
\affil{Harvard/Center for Astrophysics, Cambridge, MA}

\and 

\author{Rachael Lynn Beaton and Steven R. Majewski}
\affil{Department of Astronomy, University of Virginia, 
       P.O. Box 400325, Charlottesville, VA}

\email{courteau,widrow@astro.queensu.ca, 
 mcdonald@space.mit.edu, kgilbert@astro.washington.edu,
 raja@ucolick.org, yzhu@cfa.harvard.edu, 
 rlb9n@mail.astro.virginia.edu,
 srm4n@virginia.edu}


\begin{abstract}
We have constructed an extended composite luminosity profile for the
Andromeda galaxy, M31, and have decomposed it into three basic luminous
structural components: a bulge, a disk and a halo.  The dust-free
Spitzer/IRAC imaging and extended spatial coverage of ground-based
optical imaging and deep star counts allow us to map M31's structure
from its center to 22 kpc along the major axis.
We apply, and address the limitations of, different
decomposition methods for the 1D luminosity profiles and 2D images.
These methods include non-linear least-squares and Bayesian Monte-Carlo
Markov-chain analyses.
The basic photometric model for M31 has a \sersic bulge with shape
index $n \simeq 2.2 \pm .3$ and effective radius $R_e = 1.0 \pm 0.2$ kpc,
a dust-free exponential disk of scale length $R_d = 5.3 \pm .5$ kpc; 
the parameter errors reflect the range between different decomposition
methods.  Despite model covariances, the convergence of solutions 
based on different methods
and current data suggests a stable set of structural parameters.
The ellipticities ($\epsilon=1 - b/a$) of the bulge and of the disk
from the IRAC image are $0.37 \pm 0.03$ and $0.73 \pm 0.03$, respectively.
The bulge parameter $n$ is rather insensitive to bandpass effects
and its value (2.2) suggests a first rapid formation via mergers
followed by secular growth from the disk.  The M31 halo has a 2D power-law
index $\simeq -2.5 \pm .2$ (or -3.5 in 3D), comparable to that of the
Milky Way We find that the M31 bulge light is mostly dominant over the
range $R_{\rm min} \ltsim 1.2$ kpc.  The disk takes over in the
range 1.2 kpc $\ltsim R_{\rm min} \ltsim 9$ kpc, whereas the halo
dominates at $R_{\rm min} \gtsim 9$ kpc.  The stellar nucleus,
bulge, disk, and halo components each contribute roughly
0.05\%, 23\%, 73\% and 4\% of the total light of M31 out to
200 kpc along the minor axis.
Nominal errors for the structural parameters of the M31 bulge, disk
and halo amount to 20\%.  If M31 and the Milky Way are at all typical,
faint stellar halos should be routinely detected in galaxy
surveys reaching below $\mu_i \simeq 27$ \magarc.  We stress that
our results rely on this photometric analysis alone.  Structural parameters
may change when other fundamental constraints, such as those provided
by abundance gradients and stellar kinematics, are considered simultaneously.

\end{abstract}



\section{INTRODUCTION}\label{sec:intro}

Thanks largely to its proximity and kinship with the Milky Way, the
Andromeda galaxy has been the focus of numerous investigations
of galaxy structure.  Its visual appearance suggests the existence
of at least two components within the optical radius: a central bulge 
and a disk.  However, the very faint extended structure around M31
(Ibata \etal 2004; Guhathakurta \etal 2005, hereafter Guha05;
Irwin \etal 2005; Chapman \etal 2006; Gilbert \etal 2007;
Ibata \etal 2007; McConnachie \etal 2009; Gilbert \etal 2009b;
Tanaka \etal 2010) revealed a complex portrait of galaxy structure
that deviates considerably from the original picture of a smooth
two-component model for M31.  Furthermore, the high-resolution imaging
with the HST/WFPC1-2 also shows the existence of a small double-peaked
nucleus (Lauer \etal 1993, 1998; Kormendy \& Bender 1999, hereafter KB99).
Early measurements of M31's luminosity profile from photo-electric photometry
(de Vaucouleurs 1958), photographic plates (Walterbos \& Kennicutt 1988)
and heterogeneous digital imaging and star counts
(Pritchet \& van den Bergh 1994, hereafter PvdB94)
suggested a dominant de~Vaucouleurs (1948) profile for M31's bulge.
This impression was reinforced by the deep star counts along the
minor axis of M31 (Irwin \etal 2005, hereafter Irwin05), which yielded
brightness profiles in Johnson-$V$ and Gunn-$i$ bands below 31
and 29 \magarc, respectively.  Irwin05 suggested that a single
de Vaucouleurs profile reproduced the minor axis light of M31
within 1.4\deg\ or 19.2 kpc.  They also noted (though did not
model) that the light profile between 0.1-0.4\deg\ called for the
presence of a faint inner disk.

The rigorous decomposition of structural components in galaxies 
is critical to our understanding of their formation and evolution.  
For instance, whether a galaxy surface density profile closely
resembles a pure exponential disk (Freeman 1970) or a de~Vaucouleurs
profile is either indicative of a quiet recent history 
(e.g., Toth \& Ostriker 1992) or a turbulent past
(e.g., Bournaud \etal 2007).

In this paper, we examine the luminosity distribution of M31 based
on ground- and space-based images and explore the limitations of
the various methods that are normally used to extract structural
parameters from 1D brightness profiles and 2D galaxy images.  We
model the stellar components of M31 as the sum of a \sersic (1968)
bulge, an exponential disk and a halo.  
Despite the limitations of the different modeling procedures,
we are able to determine the structural parameters for the main
stellar components of M31 with reasonable accuracy.

We present in \se{data} the surface brightness (hereafter ``SB'')
profiles and star counts from different sources for M31.  These 
will be combined into a single, extended composite luminosity
profile for decomposition into various structural components. 
We discuss in \se{models} the various parameterizations that
are most commonly used for decomposition of the 1D and 2D light
distribution of the bulge, disk and halo components in galaxies.
Basic 1D and 2D bulge-disk decomposition methods are introduced
in \se{methods} and model results are compared in \se{BDcomp}.
The effect of a halo component is presented in \se{composite}
where we analyse the extended composite light profile of M31. 
We compare in \se{results} our results for the structural
parameters for the bulge, disk and halo of M31 with those
from the literature.  We conclude in \se{discussion} with various
interpretations of our results and reflect on future decompositions
of the M31 structure based on photometric as well as spectroscopic
information. 
Throughout this paper, we adopt a distance D$_{\rm M31} = 785 \pm 25$ kpc
(McConnachie \etal 2005). 
Thus, at M31, 1$\arcsec = 0.0038$~kpc (1$^\circ = 13.7$~kpc). 
To avoid confusion, all radii labeled ``$R$'' refer to a radius measured
along the major axis of the galaxy; radii measured along the minor axis
of the galaxy are specifically labelled as ``$R_{\rm min}$''.


\section{MEASUREMENTS OF M31'S LUMINOSITY DISTRIBUTION}\label{sec:data}

\subsection{Surface Brightness Profiles}\label{sec:SBprof}

\begin{figure*}
\begin{center}
\includegraphics[width=1.0\textwidth]{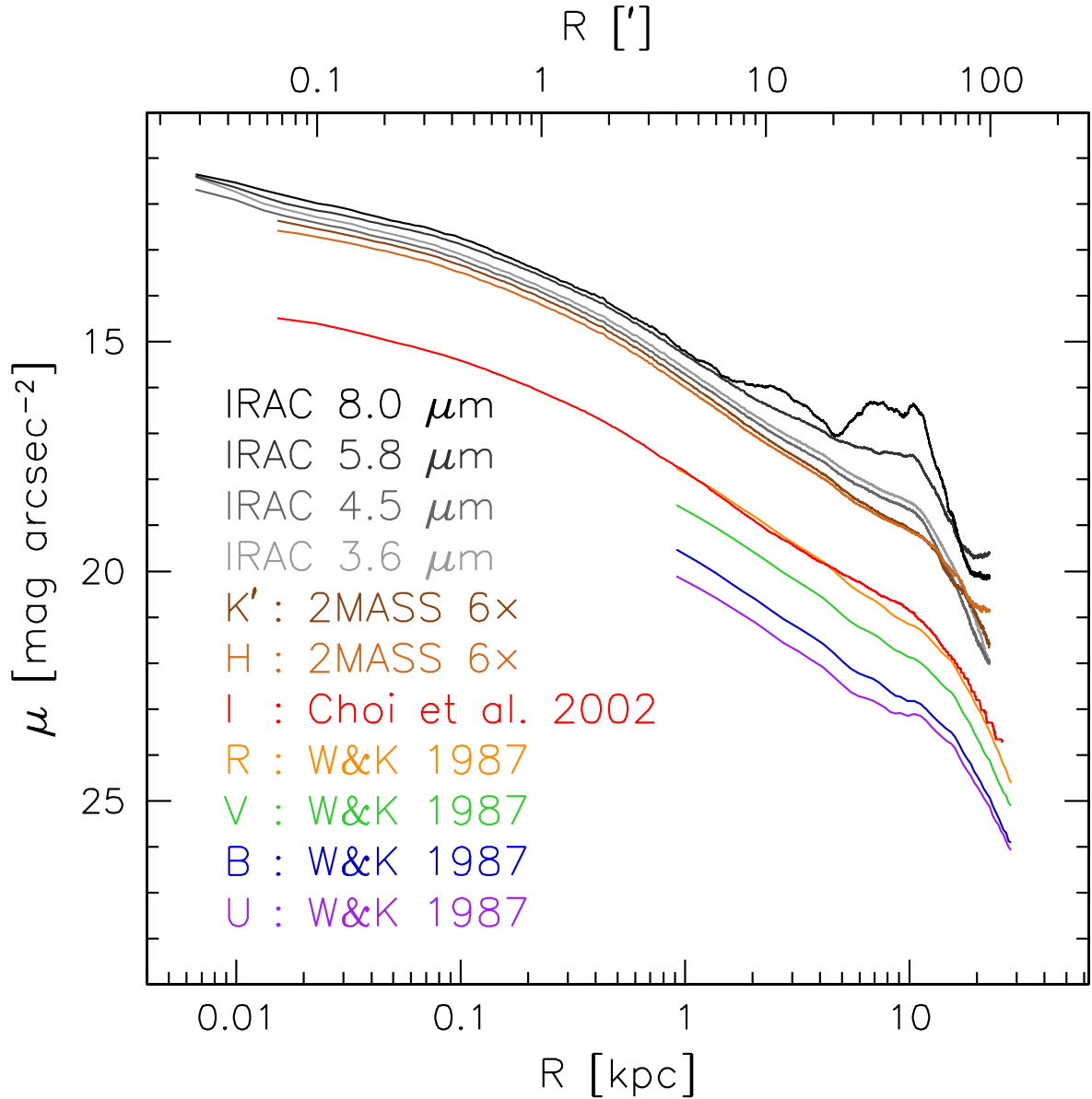}
 \caption{Azimuthally-averaged surface brightness profiles along the
   major axis for M31.  The $UBVR$ profiles were extracted from
   photographic plates by Walterbos \& Kennicutt (1987).  The
   $I$-band, 2MASS 6X $H,K^\prime$, and IRAC surface brightness
   profiles were extracted by us from the composite images of Choi
   \etal (2002), Beaton \etal (2007), and Barmby \etal (2006),
   respectively.
 }
\label{fig:M31SB}
\end{center}
\end{figure*}

The library of SB profiles for M31 is vast.
Table 1 presents a list of the major photographic and digital
data bases for luminosity profiles and star counts of M31.  
The SB profiles extracted along the major axis from the galaxy
images listed in Table 1 are shown in \Fig{M31SB} (star counts
are presented in \Fig{M31composite}). 

The much-cited library of $UBVR$ luminosity profiles of M31 from
photographic plates by Walterbos \& Kennicutt (1987) already 
highlights the so-called ``10 kpc arm'', a noticeable bump
above a rather exponential disk profile.  However, these 
optical light profiles do not sample the galaxy center well;
the starlight in the $UBV$ filters is also sensitive to dust
extinction and mostly dominated by bright massive stars, which
contribute only a small fraction of the total galaxy stellar
mass.  KB99 combined and analysed various $V$-band light
profiles of M31.  However, we focus here on redder, less
extinction-prone and more spatially-extended luminosity profiles.

Choi \etal (2002, hereafter Choi02) derived 1.7\deg $\times$ 5\deg\ 
$B$ and $I$-band CCD mosaics of M31, including M32 and NGC 205, with
the Kitt Peak National Observatory Burrell Schmidt telescope.  Each 
mosaic included 35 fields, with a typical exposure time of 40 minutes
per field in $I$.  An azimuthally-averaged SB profile of this $I$-band
mosaic was extracted via isophotal fitting by Worthey \etal (2005);
see the red line in \Fig{M31SB}.
For the purpose of the current analysis, we used the same $I$-band
mosaic from Choi02 to extract a revised $I$-band azimuthally-averaged
SB profile as well as $I$-band major- and minor-axis
wedge cuts (see \se{wedge}).
We reassessed the absolute calibration of the Choi02 data 
by performing aperture photometry of field stars in that mosaic
and adjusting the stellar magnitudes to those computed by the SDSS.
We used the same point spread function fitting routines for
the Choi02 and SDSS star fields.
This procedure yielded a photometric accuracy
of $\sim$0.1 mag (McDonald \etal 2009, hereafter McD09).  
The calibration from SDSS stars led to a readjustment of 0.4 mag
from the photometry reported in Worthey \etal 2005. 

\Fig{M31SB} shows that the Choi02 data extend as far as the
Walterbos \& Kennicutt (1987) profiles and better sample the central
regions.  For this reason, and because the rest frame $I$-band is
a better tracer of the stellar mass than $UBV$, we use the Choi02
data as our principal anchor for the derivation of a composite light
profile for M31.  While it would be desirable to combine the data
presented in Table 1 into one global composite profile, color
gradients thwart that project.  Different bands yield different
structural parameters, owing to differently distributed stellar
populations and dust, selective dust extinction, etc.
 
High-resolution, wide-field infrared mosaics of M31 remain scant due
to the smaller size and lower efficiency of IR arrays, as compared
to optical CCDs, and the significant complications due to the rapidly
varying infrared sky background.  Existing infrared maps come from the
2MASS 6X survey (Beaton \etal 2007) and the Spitzer IRAC survey (Barmby
\etal 2006).  Beaton \etal (2007) produced a 2.8 deg$^2$ JHK map of
M31 as part of the 2MASS 6X program\footnote{The 2MASS 6X program was
intended to probe about one magnitude deeper than the 2MASS
Large Galaxy Atlas (LGA) of Jarrett \etal (2003).}.  The exposure
time per source is 6 times that of the nominal 2MASS integration,
or six times 1.3s per frame times 6 frames per source: that is 46.8s
per source.  We have extracted NIR SB profiles from the 2MASS
6x JHK mosaics according to standard surface photometry
techniques (e.g., Courteau 1996; McD09).
Counts were calibrated to magnitudes using 2MASS background stars. 
SB levels were inferred from the 2MASS 6X images using
the following transformation: ${\rm SB[mag~arcsec}^{-2}\rm{]} = -2.5
\log(I{\rm[counts]}) + z.p.$, where the zero points are J:22.81,
H:22.35, and K:21.85 mag arcsec$^{-2}$.  The $H$ and $K^\prime$-band
2MASS 6X azimuthally-averaged SB profiles are reported in
\Fig{M31SB} as brown lines.  The 2MASS data (both LGA and 6X)
remain ill-constrained beyond the 10 kpc arm due to uncertain
sky subtraction (Barmby \etal 2006).  Indeed, while the proximity
of M31 enables a sharper view of its stellar content, its large
apparent size is detrimental for the reliable measurement of the sky
level far from the galaxy on timescales shorter than those of the
natural fluctuations of the IR sky.  Producing an accurately 
sky-subtracted image for M31 is an extremely challenging task
that we address in a forthcoming publication (Sick \etal, in prep.).
We therefore leave out the 2MASS data from our analysis.

The Spitzer mosaic of Barmby \etal (2006) covers 3.7\deg\ $\times$ 1.6\deg\ 
in the four IRAC bands at 3.6, 4.5, 5.8, and 8 $\mu$m.  These mosaics sample
the galaxy light with 0.861\arcsec\ pixels 
from the center out to the ``10 kpc'' arm.
The integration times per pixel ranged from 62s in the disk to 107s
in the outskirts.  Our extracted IRAC azimuthally-averaged SB
profiles are shown as grey-to-black lines in \Fig{M31SB}.
For our study, we discarded the 5.8 and 8 $\mu$m profiles which are
spoiled by PAH emission, especially near the 10 kpc arm.  The stellar
light at 3.6$\mu$m and 4.5$\mu$m bands is significantly less contaminated
by hot dust features.  For simplicity, we only used the 3.6$\mu$m
for our analysis; the 3.6 and 4.5$\mu$m data yield similar profile
decompositions.

Although the Spitzer/IRAC images do not have the superb resolution
of HST, the IRAC luminosity profiles in \Fig{M31SB} already show the
suggestive signature at R~$\ltsim 15$ pc of a nuclear component. 
We shall address the modeling of the nucleus in \S{3.1}.

It is interesting to note that the signature of the 10 kpc arm
shifts inwards as a function of wavelength.  This is expected 
if the blue stars lie at the front of the spiral density wave
where they were formed; the dust (e.g., PAHs) that results from
the evolution of those massive stars will naturally drift behind
the stellar front, as observed clearly in the brightness enhancement
between 5 and 9 kpc of the IRAC5.8/8.0 $\mu$m light profiles in \Fig{M31SB}.

\subsection{Logarithmic Wedge Profiles}\label{sec:wedge}

The luminosity profiles shown in \Fig{M31SB} are all extracted from
azimuthally-averaged isophotal fits.  However, as with most other
spiral galaxies, the light of M31 is a superposition of multiple
components with different ellipticities.  Isophotal fitting 
software packages, originally designed for single-component elliptical 
systems\footnote{On the original galaxy isophotal fitting programs,
see Young~\etal (1979), Kent (1983), Lauer (1985), Djorgovski (1985),
and Jedrzejewski (1987).},
average out the ellipticity information of a galaxy at each pixel
such that the fitted ellipses no longer represent one independent
structure.  This is perhaps the major drawback of profile decompositions 
based on azimuthally-averaged luminosity profiles.

Fortunately, for the case of M31, the position angles of the bulge
and disk are within 20\deg\ of each other, which lessens the mismatch
of superimposed isophotes.  However the ellipticities of the two components
are still quite different (see \se{results}).  Consequently, we
shall also explore 1D profile decompositions based on the simultaneous
modeling of minor and major axis wedge cuts.  Such cuts then allow
for the modeling of independent galaxian structures. 
We construct wedge cuts at different azimuthal angles as follows.
In order to keep a high signal-to-noise along the cut, 
we adopted square bins with variable width, $W$, given by: 
\be
W= N p \left( e^{n/N} -1 \right), 
\label{eq:wedge}
\ee
where $n$ is the bin number, $p$ is the minimum number of pixels at
small radius and $N$ is a constant.  The opening angle of the wedge
is thus a non-linear function of galactocentric radius.  We have used
$p=3$ and $N=25$.

\begin{figure*}
\begin{center}
\includegraphics[width=0.3\textwidth,angle=270]{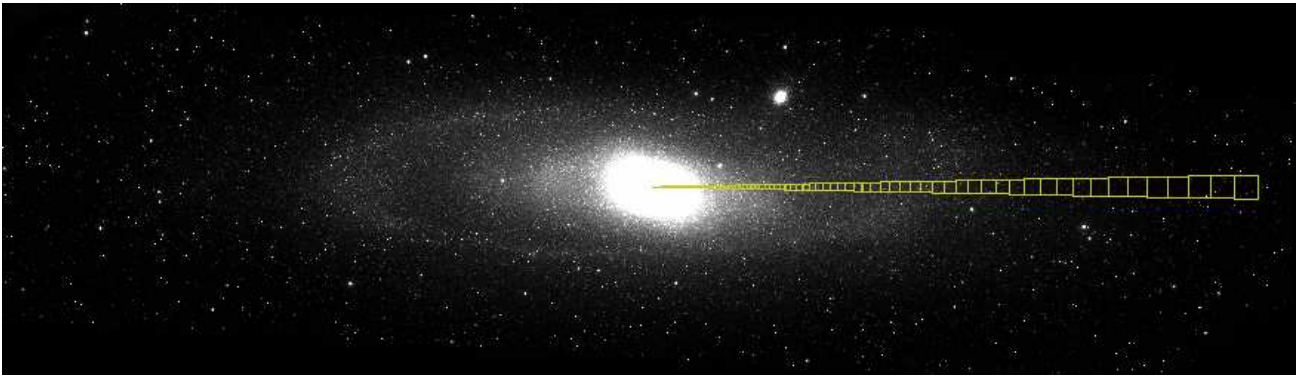}
\caption{Logarithmic wedge cut (described in \se{wedge}), 
superimposed on the IRAC $3.6\mu m$ image of M31. Each
square represents a single bin.}
\label{fig:wedge}
\end{center}
\end{figure*}

\Fig{wedge} shows the position of the major axis wedge onto the
IRAC $3.6\mu$m image.  A median SB is calculated in each variable-width bin. 
To remove bright stars, all pixel fluxes more than 3-sigma above
this value are excised and the SB profile is recomputed.  

\begin{figure*}
\begin{center}
\includegraphics[width=1.0\textwidth]{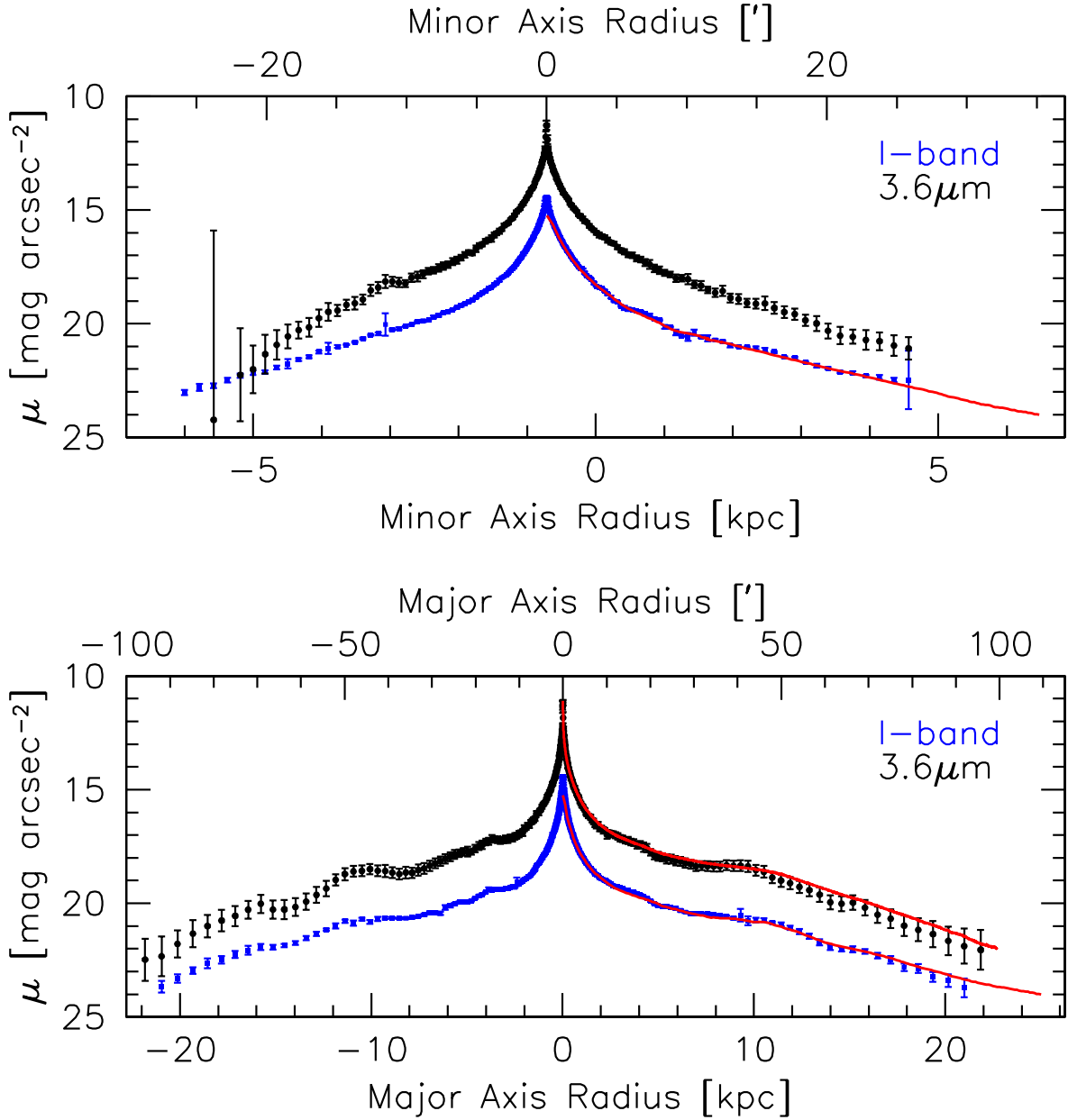}
\caption{Minor (upper panel) and major (lower panel) axis brightness
  profiles for M31, at $I$-band (blue squares) and IRAC 3.6 $\mu$m bands
  (grey and black circles), as described in \se{wedge}.
  For comparison, our azimuthally-averaged profiles for the Choi02 
  and IRAC 3.6$\mu$m data are shown in red.}
\label{fig:M31cuts}
\end{center}
\end{figure*}

The minor and major axis wedge cuts extracted from the Choi02 
and IRAC $3.6\mu$m images are shown in \Fig{M31cuts}.  The major 
axis cuts are also compared with the azimuthally averaged SB
profiles at $I$-band and at 3.6$\mu$m. Despite our concerns that azimuthal
averaging yields, in principle, an unrepresentative blend of distinct
luminous components each with different ellipticities, the
azimuthally-averaged profiles for M31 differ only slightly from the
precise cuts.  The cuts and averaged profiles are thus interchangeable
but the latter have higher signal-to-noise levels.  

\begin{figure*}
\begin{center}
\includegraphics[width=1.0\textwidth]{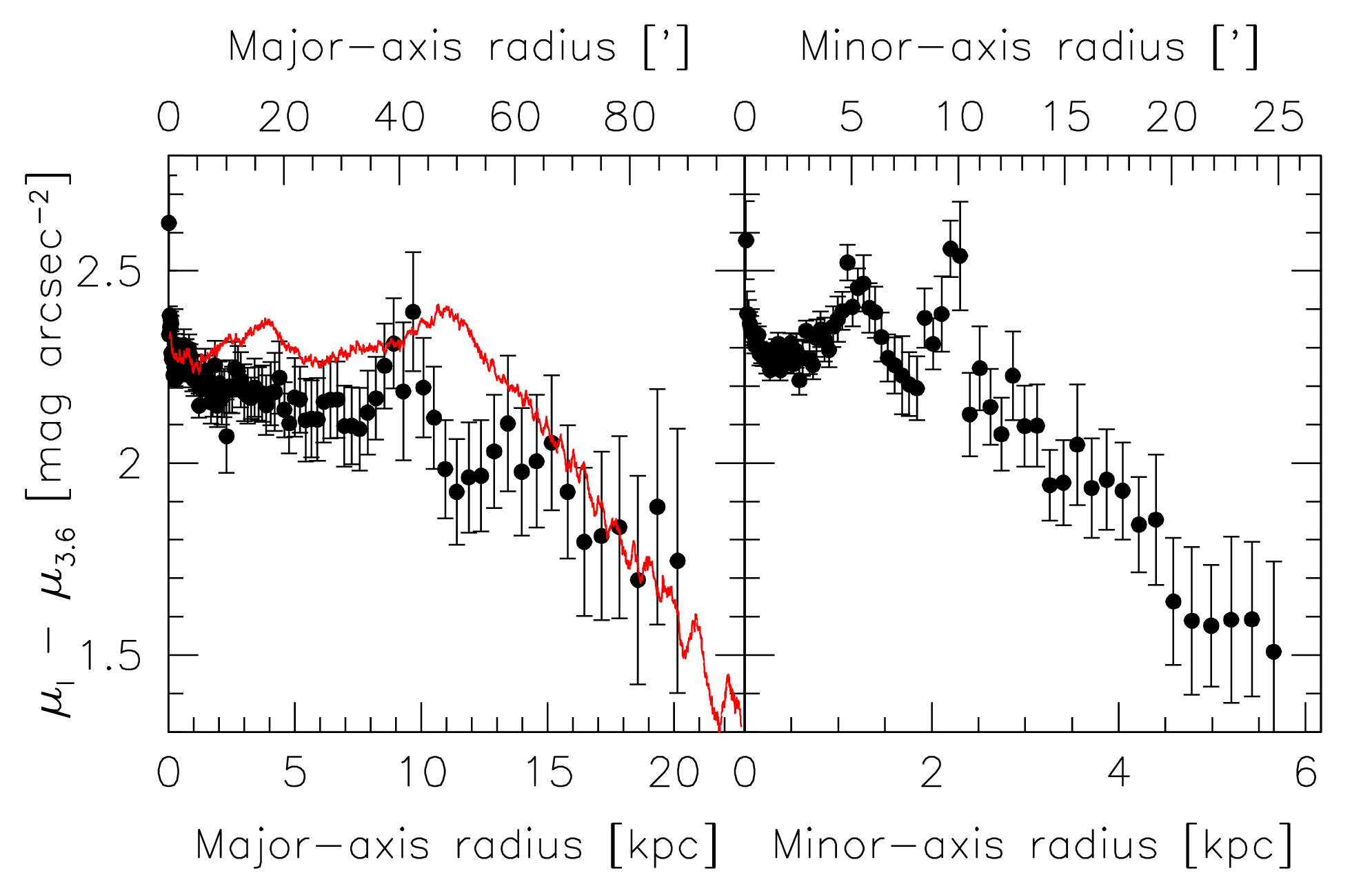}
\caption{
I-3.6$\mu$m color gradients from wedge cuts along the major (left)
and minor (right) axes of M31.  The color gradient from the azimuthally-averaged 
surface brightness profile (projected along the major axis) is shown as
the red line on the left side.  The strong gradient ($\sim$0.5 \magarc\
over the 90$^\prime$ extent of the major axis) is a clear indication 
that the $I$-band and IRAC luminosity profiles cannot be merged without
radially-dependent color corrections.  Note the clear bulge-disk
transition at roughly 8-10 kpc.  Both the M31 bulge and disk get
bluer with radius, though at different rates. 
}
\label{fig:M31colgrad}
\end{center}
\end{figure*}

Subtraction of the $I$ and $3.6\mu$ cuts in \Fig{M31cuts} reveals
color gradients in the inner parts ($R_{maj}<20$ kpc) of M31. 
These gradients, shown in \Fig{M31colgrad}, are consistent 
with the bulge being generally populated by more metal-rich 
stars than the disk, as is seen in other Sb-Sc like galaxies
(MacArthur \etal 2004; Moorthy \& Holtzman 2006; MacArthur etal 2009;
 Roediger \etal 2011). 
The blueing of the disk, especially beyond the 10 kpc ring, 
confirms that the $I$-band and 3.6 $\mu$m luminosity profiles cannot
simply be merged (without corrections) to build an extended
composite light profile for M31.

\subsection{Profile Error Bars}\label{sec:errors}

Rigorous error bars are crucial to the proper model decomposition of
a luminosity profile.  For azimuthally-averaged SB profiles,
the quoted intensity per radial increment (here, per pixel) is the median
of all the pixel intensities along a given elliptical isophote (light
contour) and the brightness error per radial bin is the standard deviation
of all those pixel intensities along the isophote with respect to the
median intensity value (Courteau 1996).  For a wedge cut along any
radius, the median intensity per bin is computed as in \se{wedge};
the quoted error is the rms deviation about the median value
of the pixels in each bin.  The latter error estimates
can never be an exact representation of the true galaxian (sky-subtracted)
photon noise at each bin since each pixel element is smaller than the
seeing disk, and all adjacent pixels are thus correlated.  The size
of the error bars per bin is however comparable to the amplitude of the
point-by-point fluctuations (see \Fig{M31cuts}) suggesting that our
error estimates are reasonable. 

The error bars as quoted by the authors for the Irwin05 profiles
are significantly larger, at large radii, than the point-by-point
fluctuations.  For our modeling of the Irwin05 data, we have recomputed
those errors as the standard deviation about linear fits through
neighboring data points.  These errors closely reflect the
point-by-point fluctuations of the data. 

\subsection{Star Counts}\label{sec:counts}

In addition to the galaxy images and surface brightness profiles,
information about the shape of the galaxy components can be gleaned
from star counts measured in fields around the disk of the galaxy
where crowding effects are lessened.  
We have combined the extended star counts of the M31 stellar halo
largely along the minor axis by Pritchet \& van den Bergh (1994,
hereafter PvdB94), Irwin05, and Gilbert \etal (2009a, hereafter
G09).  These star counts cover the range
20 kpc $\leq R_{\rm min} \leq$ 150 kpc (along the minor axis;
see also Ibata \etal 2007 and McConnachie \etal 2009). 

\begin{figure*}
\begin{center}
\includegraphics[width=1.0\textwidth]{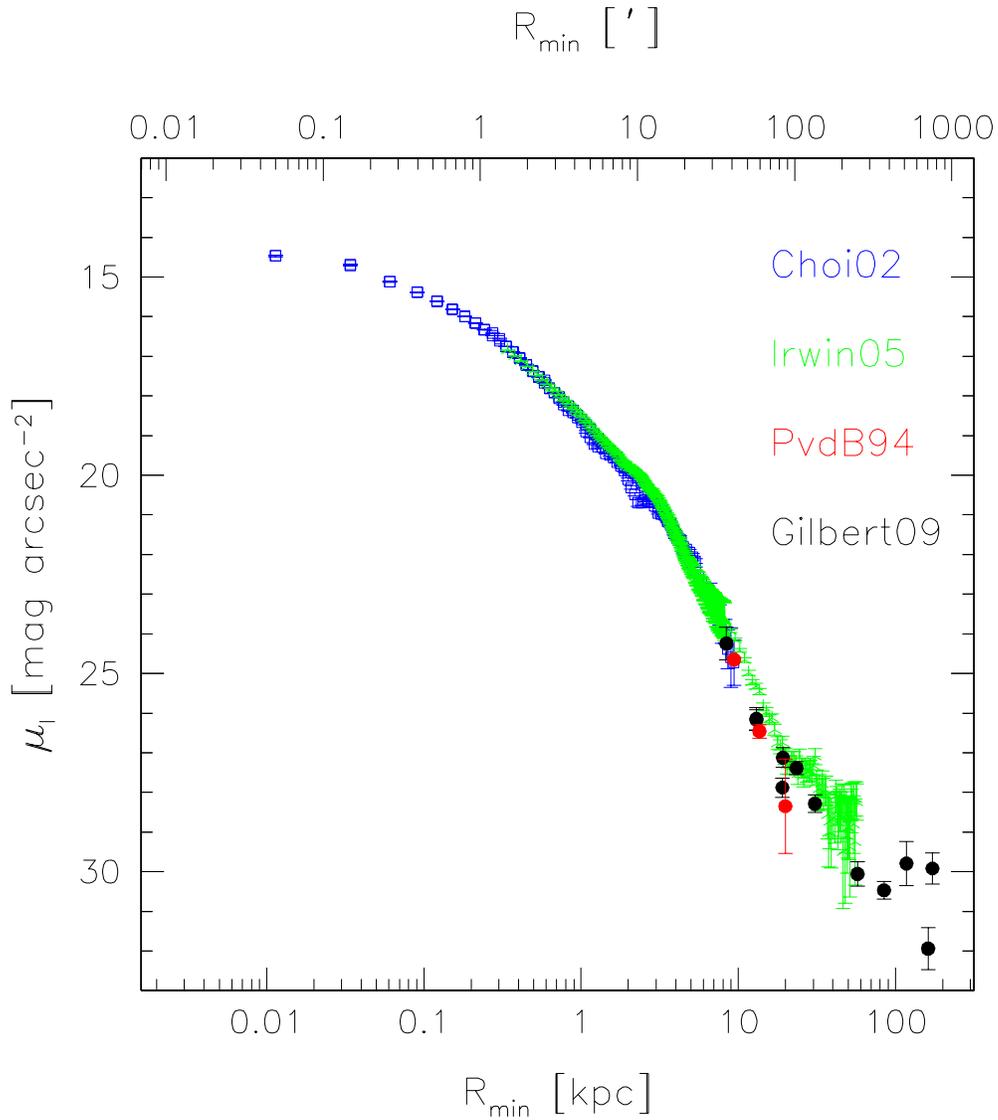}
\caption{Composite minor-axis profile, described in \se{counts}.
  The radius is projected along the minor-axis. 
  The $I$-band minor-axis brightnesses, shown in blue, were extracted
  from the $I$-band image of Choi \etal (2002 [Choi02]). Extension of 
  the Choi02 profile relies on the star counts of the M31 stellar halo
  largely along the minor axis by Irwin \etal (2005 [Irwin05] in 
  green, Pritchet \& van den Bergh (1994 [PvdB94) in red and 
  Gilbert \etal (2009a) [Gilbert09] shown in black.  The Irwin05
  error bars were rederived by us (\se{errors}).}
\label{fig:M31composite}
\end{center}
\end{figure*}

PvdB94 obtained digital star counts with the Canada-France-Hawaii
Telescope in the $V$-band for 8 fields along M31's minor axis.
Each field was observed three times for a minimum of 900s per
integration. These counts were then combined with photographic
luminosity profiles by de~Vaucouleurs (1958) and Walterbos \& Kennicutt (1987)
and digital imaging by Kent (1983).  The PvdB94 minor-axis 
brightness profile is shown with red circles in \Fig{M31composite}. 
Based on these data, PvdB94 postulated that the halo of M31
along the minor axis was well-represented by a single de~Vaucouleurs
law over the range 0.2 kpc $\leq R_{\rm min} \leq$ 20 kpc.

Irwin05 used stellar counts from the Isaac Newton Telescope to
determine M31's luminosity profile along its south-east minor axis.
They combined PvdB94's data with faint red giant branch star
counts to trace the minor axis stellar distribution out to a
projected radius of $\sim$ 55 kpc, where $\mu_V \sim$ 32 mag arcsec$^{-2}$.
They exposed typically 800-1000s per field per bandpass in
Johnson $V$ and Gunn $i$.  We show their Gunn $i$ minor-axis
luminosity profile as green dots in \Fig{M31composite}.  Irwin05's analysis
corroborated PvdB94 with a de~Vaucouleurs profile out to a projected
minor axis radius of $\sim 20$ kpc.  Beyond that radius, 
Irwin05 surmised, the light profile would assume a more exponential
shape with a scale length of 14 kpc.  We will see in \se{composite}
that the M31 minor-axis cut is indeed a poor tracer of disk light.
Indeed, our decompositions argue against a de~Vaucouleurs profile
(\sersic $n=4$) for the M31 bulge.

Guha05 presented a $V$-band SB profile
along M31's south-eastern minor axis based on counts of resolved
member red giant stars in a Keck/DEIMOS spectroscopic survey.
The spectroscopic exposure time with Keck/DEIMOS was 1 hour per field. 
Red giants in M31 were identified (and distinguished from foreground Milky
Way dwarf star contaminants) using a combination of photometric and
spectroscopic diagnostics (Guhathakurta \etal 2006; Gilbert \etal 2006). 
To estimate M31's stellar surface density as a function of radius, 
the observed ratio of M31 red giants to Milky Way dwarf stars
was multiplied by the surface density of Milky Way dwarf stars
predicted by the ``Besan{\c c}on'' Galactic star-count model 
(e.g., Robin \etal 2003; Robin \etal 2004). 
This estimate was then calibrated to $V$-band intensity by comparing with the
PvdB94 data in overlapping regions (also based on star counts).
The minor-axis SB profile from $V$-band counts,
presented here as black dots in \Fig{M31composite}, includes two
improvements relative to the Guha05 analysis -- as reported 
in G09: 
(1)~ a larger number and wider spatial distribution of Keck/DEIMOS 
spectroscopic fields providing denser sampling over the radial range
$10 < R < 165$~kpc (Kalirai \etal 2006; Gilbert \etal 2007), and 
(2)~correction for field-to-field variations in the spectroscopic
sample selection function.

\subsection{Composite Minor Axis Profile}\label{sec:composite}

The extended minor axis composite profile of M31 shown in \Fig{M31composite}
is a combination of the minor axis wedge cut from the image
of Choi02 (\se{wedge}) and the deep minor axis star counts of Irwin05,
PvdB94, and G09.  The star counts were all normalised
to our recalibrated $I$-band profile for Choi02.  Altogether,
our composite profile yields a detailed picture of M31's global
$I$-band luminosity distribution for $\mu_I < 30$ \magarc. 
\Fig{M31composite} may be compared, for $R_{minor} < 80$~kpc, 
with Fig.~31 of Ibata \etal (2007), using $V-I \sim 2$.

Our modeling of the composite profile can either treat 
all the data points independently (``Unbinned''), or use
a uniformly-sampled profile that has been averaged over all
overlapping data (``Binned''). Both techniques are often
found in the literature and we wish to test them here.
A weighted fit should in principle yield similar solutions
for both the unbinned and the binned profiles since the
former will have more data points per bin (in regions
of data overlap) while the latter will have, on 
average, smaller error bars per bin.  Reality is not 
always so simple. 

In order to create a single, composite, ``binned'' minor axis
profile from the four above-mentioned profiles, we adopted a
logarithmic binning as in \Eq{wedge} with $p=3$ and $N=25$.
These parameters were chosen to match the inner regions of the 
$I$-band profile minor axis profile.  The surface brightness
per bin is the error-weighted average of the surface brightnesses
from all of the available data in that bin.  The error per bin
is the root-mean-square of all the data point errors in that bin.
We will analyse both the unbinned and binned composite profiles
in \se{results}. 


\section {GALAXY MODEL COMPONENTS}\label{sec:models}

We decompose the M31 1D luminosity distribution and 2D
image into basic galaxy model components.  Visual examination
of \Fig{M31SB} and \Fig{M31composite} suggests the existence
of at least four distinct stellar components in M31: a nucleus,
a bulge, a disk, and a halo.  

The 1D luminosity profiles for our bulge and disk (hereafter B/D)
decompositions without a halo component consist of the $I$-band
and IRAC 3.6 $\mu$m azimuthally-averaged (hereafter 'AZAV') SB
profiles (\Fig{M31SB}) and major- and minor-axis cuts (\Fig{M31cuts}).
As seen in those Figures, those profiles extend out to roughly 6 kpc
along the minor axis, or 23 kpc along the major axis.
We use the $I$-band composite minor-axis profile (\Fig{M31composite}), 
which now extends to out to 200 kpc along the minor axis, for
decompositions that include a halo component.  We also use the IRAC
3.6 $\mu$m image to derive a model decomposition of the 2D light
distribution within 23 kpc using the GALFIT program of Peng \etal 
(2002, hereafter GALFIT).  The IRAC image is preferred for
2D modeling to that of Choi02 for its higher resolution,
lesser sensitivity to dust, and comparable spatial extent. 
However, given its relatively shallow depth, the IRAC image
of M31 cannot be used to model the halo component.

We now discuss the galaxy model component parameterizations. 
Except for the nucleus, seeing convolution of the model 
functions is unnecessary for the nearby M31.

\subsection{Galaxy Nucleus}\label{sec:nucleus}

As noted above, Lauer \etal (1993) were first to show with
their HST/WFPC1 images that the M31 nucleus was double-peaked. 
Despite the poorer resolution of our IRAC images, and our
inability to resolve two distinct sources, the gentle 
central rise of the IRAC light profiles for $R < 15$ pc
in \Fig{M31SB} is clearly suggestive of a nucleus component.  
KB99 used Lauer's HST/WFPC1-2 images of the M31 core and high
spatial-resolution CFHT/SIS spectra to infer that the M31 nucleus
is a separate component with an origin different from that
of the bulge.  The well-resolved HST $V$-band images analysed
by KB99 show a nearly exponential nucleus (see \Eq{exp} below)
with a scale length $R_n=0.88$\arcsec$=3.3$pc (Kormendy, private comm.). 
The nucleus contributes only a very small fraction ($< 0.05\%$)
of M31's total luminosity budget.  M31's nucleus is indeed
sufficiently small that the model parameters for the bulge,
disk and halo are largely unaffected by it.
Therefore, while we embrace the results by Lauer \etal (1993)
and KB99, we now ignore the nucleus component in our luminosity
profile decompositions.

\subsection{Galaxy Bulge}

We assume that the 2D luminosity distribution of the M31 bulge
has a projected profile given by the \sersic (1968) profile,
\be
\label{eq:sersic}
I_b(R)=I_e\exp\left\{-b_n\left[\left({R\over{R_e}}\right)^{1/n}-1\right]\right\}, 
\ee
where $R_e$ is the projected half-light radius, $I_e$ is the
intensity at $R_e$, the exponent $n$ is the \sersic shape
parameter, and $b_n = 1.9992n - 0.3271$ (Capaccioli 1989).
With $n=1$ or $n=4$, the \sersic function reduces to the
exponential or de~Vaucouleurs functions, respectively. 

The total extrapolated luminosity of a \sersic\ bulge is given by:

\be
\label{eq:totalbulge} 
L_b=2\pi (b/a) \int_{0}^{\infty}I_{b}(R) R\,dR=2\pi (b/a) I_{e}R_{e}^{2}{{e^{b_{n} 
\,}n\Gamma(2n)}\over{b_{n}^{2n}}}
\ee
where $a$ and $b$ are the semi-major and semi-minor axes of the bulge.
We will use \Eq{totalbulge} to compute the bulge total light fraction
in \se{results}.  See Graham and Driver (2005) for more details about
the \sersic function.

\subsection{Galaxy Disk} 

We assume that the projected light profile of the disk is described
by a basic exponential function,
\be
\label{eq:exp}
I_d(R)=I_0\,\exp\left\{- {R / R_d } \right\}, 
\ee
where $I_0$ is the disk central intensity and $R_d$ is the 
projected disk scale length.
The total extrapolated luminosity of an exponential disk is given by: 

\be
\label{eq:totaldisk} 
L_d=2\pi (b/a) I_{0} {R_d}^{2} 
\ee
where $a$ and $b$ are now the semi-major and semi-minor axes of the disk. 

Various authors (Kormendy 1977; Baggett \etal 1998; Puglielli \etal 2010)
have considered an inner disk truncation as an alternative to modeling galaxy
light profiles with a dip at the bulge-disk transition.  Such a dip can
be understood through various effects, such as dust extinction (Mac03)
or the mixing action of a dynamical hot stellar bulge or bar
(Puglielli \etal 2010).
However, we ignore the possibility of a core in the inner disk of M31,
as the current study based on imaging data alone offers no leverage
on testing this hypothesis.
We limit our modeling of the disk to a pure exponential function with
a constant scale length at all radii.

\subsection{Galaxy Halo} 

The deep star counts from Irwin05 and G09, shown in
\Fig{M31composite}, are clearly indicative of an additional
component at $R_{\rm min} \gtsim 12$ kpc, which appears to be independent
of the galaxy disk (see also Ibata \etal 2007 and Tanaka \etal 2010).  
As shown by Gilbert \etal (2007; 2009), the kinematics of the
stars in tidal streams out to 60~kpc show a large spread in velocities,
consistent with expectations for a halo population.  
In \se{halo}, we also suggest that the M31 bulge and halo components
are likely independent on account of their different spatial distributions.

We model the (1D) halo with either a \sersic function (\Eq{sersic})
or a power-law: 
\be
I_{\rm h}(R) = I_* \left\{ \frac{1 + \left (R_*/a_h\right)^2} 
{1 + \left (R/a_h\right )^2}\right\}^\alpha 
\label{eq:faint}
\ee
where $R_*$ is a turnover radius and $I(R_*) = I_*$.
The magnitude formulation is simply $\mu_{h} = -2.5 \log I_{h}$. 
Based on visual examination of \Fig{M31composite}, we adopt
$R_* = 30 \,{\rm kpc}$ (measured along the minor axis) where the 
halo is dominant.  The value of $R_*$ is arbitrary and only affects $I_*$.
The quantity $a_h$ sets the amplitude of the profile in the inner parts.  
Note that at large radii, the Hernquist model (Hernquist 1990)
and NFW function (Navarro \etal 1996), as explored for the M31
halo by e.g., Ibata \etal (2007) and Tanaka \etal (2010), are
special cases of our power-law model.

For the power-law halo, the total luminosity is
\be
\label{eq:totalhalo}  
L_h= { 2\pi (b/a) I_{*} {R_*}^{2} \frac{1 + a_*^2}{2\left (\alpha-1\right )} 
\left \{
\left (\frac{1 + a_*^2}{a_*^2}\right )^{\alpha-1} - 
\left (\frac{1 + a_*^2}{s_{\rm max}^2 + a_*^2}\right )^{\alpha-1}
\right \} } 
\ee 
where $a_* = a_h/R_*$ and $s_{\rm max} = R_{\rm max} / R_*$. 
In this work, we integrate out to a maximum radius 
$R_{\rm max} = 200$ kpc along the minor axis. 

We also assume a spherical halo ($a=b$) as the current star counts
offer no leverage on the halo shape.  If M31's halo is anything like
the Milky Way's, our assumption of a spherical halo is well justified
(Majewski \etal 2003; De Propris \etal 2010; although see Bell \etal 2008).


\section{DECOMPOSITION METHODS}\label{sec:methods}

To achieve model decompositions of the galaxy's 1D light
profiles (azimuthal averages or radial cuts) and 2D image,
we use different basic fitting methods based on frequentist
or Bayesian methodologies. The 1D azimuthally-averaged SB profiles
and wedge cuts for M31 can be decomposed into the different 
model components above using a non-linear least-squares (hereafter
``NLLS'') minimization method based on the Levenberg-Marquardt
algorithm (downhill gradient).
Such a ``frequentist'' approach for the decomposition
of galaxy components has been implemented by many
(e.g., Kent 1985; KB99; Mac03; McD09).
The robustness of the ``NLLS'' method is discussed in Mac03.
The model parameter one-sigma error bars are computed as
in Mac03 and McD09.  Those errors reflect
the 68\% range of solutions from a full suite of
Monte Carlo NLLS realisations with a complete range
of initial guesses over all possible fitted parameters.
Because the adopted structural models are usually 
not ideal representations of the real structure of
a galaxy, the statistical errors per model parameter
from NLLS fitting codes are usually rather small;
much more so than the errors computed through the
Monte Carlo method alluded to above. 

An alternative approach is to implement Bayesian statistics
and a Markov Chain Monte-Carlo (hereafter ``MCMC'') analysis. 
The aim here is to determine the
posterior probability, $\pmd$, that is, the probability of the
model, M, given the data, D, and prior information, I.  Since our
model is expressed in terms of a set of parameters, P(M|D) takes the
form of a probability distribution function (PDF) over the model
parameter space.  The calculation is performed via Bayes' theorem
which states that $\pmd \propto \prm\pdm$ where $\pdm$ is the
probability of the data given the model and $\prm$ is the prior
probability on the model parameters.  In fact, $\pdm$ is the
likelihood function of the NLLS method described above.  The Bayesian
approach is more ambitious.  With NLLS, one is simply interested in
the values of the parameters that maximize the likelihood function.
Here, one must calculate (or at least estimate) $\pmd$ as defined
over the entire model parameter space.

In principle, Bayesian Inference provides a more complete
understanding of how well the model describes the data (the state of
our knowledge).  For example, one can calculate the PDF for any set of
quantities (say a subset of the model parameters or quantities derived
from the model parameters) by integrating $\pmd$ over the parameter
space (marginalization).  In this way, one can obtain confidence
limits on the parameters.  By contrast, with NLLS, confidence limits
on the parameters are obtained from the shape of the likelihood
function near the peak.  Now if the likelihood function is
single-peaked and well-behaved, and if our prior, P(M), is relatively
constant over the range where $\pmd$ is appreciable, then the peak
in $\pmd$ will be very close to the parameter values obtained by
simply maximizing $\pdm$ (though again, we stress that the calculation
and interpretation of the confidence limits is different for the two
methods).  For a review of Bayesian Inference and how it compares
with the frequentist approach, see Gregory (2005). 

Calculation of P(M|D) often involves complicated, multi-dimensional
integrals.  For example, in the case where we include a bulge, disk,
and halo component, and consider both major and minor axis profiles
(and hence, require ellipticities for the three components) the model
is described by ten parameters making a direct calculation (or, for
that matter, a grid search in a maximum likelihood scheme) prohibitively
time consuming.  Hence, we resort to the powerful MCMC method.  With MCMC, 
$\pmd$ is approximated by a chain of points in the parameter space
which is constructed via a simple set of rules.  In our work, we
employ the Metropolis-Hastings algorithm (Metropolis \etal 1953; 
Hastings 1970). 
For a chain of sufficient length, the
distribution of points along the chain {\it is} the probability
distribution function, P(M|D).  PDFs for any subset of parameters are
obtained by simply projecting the chain on to the appropriate
subspace.

The Metropolis-Hastings algorithm is described in numerous documents. 
The engine of the algorithm is the jumping rule which must be
chosen with care in order to insure that the Markov chain adequately
approximates $\pmd$.  In this paper, we use an iterative scheme 
which is similar to simulated annealing (see Puglielli \etal 2010). 

For the present analysis we choose a uniform prior for the scale-free
parameters (e.g., $\mu_d,\, \epsilon$) and a logarithmic, or Jeffrey's
prior (equal probability per decade), for the scaling parameters 
(e.g., $R_d,\, R_e$). 
In addition, we introduce a "noise" parameter for each data set.
That is, the likelihood function, $\pdm$, is taken to be 
\begin{equation}
\pdm = \prod_{ij}\frac{1}{\sqrt{2\pi\sigma_{ij}}}
e^{-\left (d_{ij} - m_{ij}\right )^2/\sigma_{ij}^2}
\end{equation}
where $\sigma_{ij}^2 = e_{ij}^2 + f_j^2$.  In these expressions, $j$ denotes
the different data sets (IRAC, Choi02, etc.) while $i$ labels the data points
within each data set.  The noise parameters, $f_j$ can account for features
in the data not explained by the model or for the possibility that the quoted
measurement errors, $e_{ij}$ were underestimated.  The noise parameters are
included in the model parameter space with a Jeffrey's prior.
For the calculation, we treat all quantities in counts.
However, the noise parameter is taken to be a fractional
error, i.e., $f_j = s_j d_{ij}$ so the actual parameter
in the MCMC analysis is $s_j$. 

We can also decompose 2D galaxy images using similar
parameterizations as for the 1D case, through the simultaneous
matching of the image pixel distribution by a 2D model, using
a downhill gradient algorithm.  Much like the simultaneous
analysis of 1D profile cuts taken at different position angles
(hereafter PAs),
the 2D multi-component fitting enables the determination of
distinct ellipticities for the underlying galaxy components.
Unlike our 1D isophotal maps, the 2D models assume a constant
ellipticity and PA for each axisymmetric component.
For the fitting of 2D parameterized, axisymmetric, functions 
to astronomical images, we here adopt the (frequentist) GALFIT
galaxy/point source fitting algorithm by Peng (2002).

To assess the robustness of the 2D fits and the sensitivity of
the final 2D galaxy model to the initial guesses, we have performed
a suite of decompositions with initial guesses spanning a wide
range of values. 
The initial guesses for the geometrical parameters
(PA and axial ratio) were set according to the
best 1D isophotal fits.  We find the 2D decompositions
to be rather insensitive to deviations of the initial guesses,
with final best-fit parameters differing by no more than 0.1\%.
Much like the 1D NLLS error analysis, the one-sigma error estimate
per fitted parameter reflects the small range of GALFIT solutions
for a broad range of initial guesses.

We have also tested for the effect of sky uncertainties.
In the 1D analysis, using the Choi et~al image, we computed
model parameters based on sky levels that differ by one sigma
from the mean.  That sigma was determined from the standard
deviation of sky levels measured in five sky boxes adjacent
to the galaxy.  With those different sky levels, the scale
radii and bulge $n$ can vary by no more than 7-9\% and the
effective surface brightnesses by less than one tenth of
a magnitude.  Similar results apply for our 2D image.  For
the latter, we have also tested for a floating sky in GALFIT.
This results in a 0.6\% decrease in the bulge scale length
and 1.3\% decrease in the disk scale length.  The geometry
of the bulge/disk system is unchanged.  All cases considered,
our results are robust again sky level uncertainties. 

For the 2D fits, we only model the IRAC 3.6$\mu$m image with
GALFIT (for reasons explained in \se{models}).  Also, the
latter requires an exposure time map, for the construction
of a pixel-to-pixel variance map, which cannot be reconstructed
for the Choi02 mosaic.  This variance map assumes Poisson
statistics, calculated from the coverage maps of the IRAC mosaics.
Since the IRAC images are too shallow to reveal the stellar halo
component, our GALFIT modeling is limited to a \sersic bulge
and an exponential disk.  Each component has its own ellipticity
and PA.  The latter is a significant advantage over our
implementations of 1D NLLS and MCMC techniques\footnote{Our
MCMC code can assess independent ellipticities but not variable
PAs; the latter would however be possible if galaxy cuts at
multiple PAs were analysed simultaneously.}.

A further complication for GALFIT errors is the issue of SB
fluctuations in the outer disk which, for M31, can be large
compared to the Poisson noise (owing to the proximity of M31).
We have tested for this effect in our construction of the
variance map for the 2D fit.  Accounting for SB fluctuation
errors can alter model parameters by $\sim 10\%$.  Those are
the typical errors that we quote for the model parameters in
Table 3.  These could be larger on account of other systematic
effects that we have not accounted for such as the modeling
of a bar and the effect of a variable tilt in the M31 disk. 
The random errors are otherwise negligible in light of the
very high signal-to-noise of the IRAC 3.6$\mu$m pixels. 

The 1D and 2D approaches to modeling galaxy luminosity profiles
and images may yield different results due to their intrinsic
differences (e.g., different convergence algorithms, weighting
schemes, accounting for PAs and ellipticities, etc.\footnote{For
discussions about the pros and cons of 1D vs 2D light modeling
techniques, see e.g., Byun \& Freeman (1995), de Jong (1996),
Peng (2002) and Mac03.}).
As we discussed in \se{models}, the choice of minimization
against counts or magnitudes can potentially yield different
solutions. 
However, we have verified that provided the errors are correctly
propagated and given the present data for M31, our minimizations
against magnitudes or counts yield similar results.


\section{COMPARISON OF METHODS}\label{sec:BDcomp}

We first address
intrinsic differences between the 1D NLLS and MCMC modeling techniques, 
as well as against 2D modeling, with a simplified data set and a simple
fitting model.  To this end, we model only a \sersic bulge and a disk
with the IRAC 3.6$\mu$m and the Choi02 data.
The B/D decompositions of the 1D IRAC and Choi02 luminosity profiles 
from both our NLLS and MCMC methods are reported in Table 2.
Decomposition parameters of the 2D IRAC image with GALFIT
are presented in Table 3.  The bulge and disk parameters 
in Tables 2 and 3 correspond to the parameters in Eqs. 2 \& 4. 
The surface brightnesses, computed as $\mu = -2.5 \log I$ \magarc,
are not corrected for projection or extinction effects.  The bulge
and disk ellipticities, $\epsilon_{bulge}=1-(b/a)_{\rm bulge}$
and $\epsilon_{disk}=1-(b/a)_{\rm disk}$, are presented where available. 

In Table 2, the column ``Cut'' can have ``Min'' and ``Maj'' for
independent minor axis or major axis cuts, respectively (see Models
A-D and I-L).  The notation ``MinMaj'' identifies the simultaneous
use of both the minor and major profile cuts with our MCMC code to
yield, like GALFIT with a 2D image, independent estimates of the
bulge and disk mean ellipticities (see Models E \& M).
``AZAV'' and ``AZAVmsk'' refer to the azimuthally-averaged 
surface brightness profile, whether raw or with spiral arms
clipping respectively (Models F-H and N).  Spiral arms clipping for
surface brightness profiles is discussed in McD09; it is meant
to exclude the portion of the profile that is affected by a spiral arm, 
revealing the underlying exponential stellar profile (an alternative
approach is also to fit, rather than clip, the spiral arm features. 
This approach introduces additional parameter covariance). 
A natural consequence of azimuthal averaging is that (non-circular)
arms appear broader in AZAV profiles than they do along any
azimuthal cut.  A thin, logarithmic spiral arm can project into
a wide azimuthally-averaged feature in a 1D AZAV 
profile, giving the illusion of a broader arm (or bar).  
Spiral arm clipping thus removes more bona fide old disk light
in 1D than it would in 2D with a non-axisymmetric model of the arms.  

Note that the structural parameters listed in Table 2 for ``Min''
cuts are those projected along the minor axis of M31, while those
listed for the ``Maj'', ``MinMaj'', or ``AZAV'' profiles refer
to projection along the major axis of M31.

While our NLLS algorithm was not coded to extract ellipticities
from cuts of different PAs, the isophotal fits of the IRAC
and Choi02 images yield ellipticity profiles that can be
compared to the MCMC values for $\epsilon_{bulge}$ and 
$\epsilon_{disk}$.  The disk ellipticities from both isophotal
fitting and the MCMC analysis are identical; these are not 
affected by the presence of a bulge.  
The bulge ellipticity, $\epsilon_{bulge}$, cannot
be estimated from isophotal fits since the inner
ellipticities are the result of superimposed
structural components.  We return to the comparison
of 1D and 2D ellipticities in \se{results}.

Given the same data and fitted models, as in Models A-D, F-G
and I-L, the NLLS and MCMC methods are nearly equivalent; 
some differences exist (e.g., Models F and G) largely due
to operational differences between the NLLS and MCMC methods.
Models F and G treat AZAV data that are far better sampled 
in the outer disk than the Min/Maj cuts for Models A-D and
I-L.  The sensitivity to the 10 kpc arm is thus enhanced in
the AZAV profile, via smaller error bars per point.
In Model G, the Bayesian MCMC analysis includes a "noise"
parameter which is added, in quadrature, to the calculated
error.  This parameter effectively softens the impact of
non-exponential spiral disk features.  The net result is
a fit that does worse, relative to F, in the center but
slightly better at large radii.
Effectively MCMC ignores the ``10 kpc'' arm signature in
the SB profile while NLLS (as coded) must force a fit.
This is made even more evident when the spiral arms are
masked in Model H; the disk structural parameters come
to a closer agreement with the MCMC Model G.  The
conservative conclusion for Model G is that it yields
a lower limit on the range of allowed values of $n$ for
the M31 bulge.  

The 1D B/D decompositions from Table 2 (Models A-H) can be
compared with our GALFIT B/D analysis of the 2D IRAC 3.6$\mu$m
image.  Results from the latter are reported in Table 3.
We have explored the effect of spiral arm clipping (Model O
versus P) but the differences are minimal.  This result may
seem surprising in light of the contrast between Models F \& H,
but recall that the spiral arms are less dominant in the 2D image.
Unlike in 1D, the 2D fitting technique primarily picks out and fits
the axisymmetric features in the data, and thus the non-axisymmetric
spiral arms have less influence in 2D than in 1D.
Indeed, unlike the variations between the NLLS Models F and G,
the comparison of GALFIT Models O and P here shows little
difference with arm masking.

We also explore in Model Q, which is unmasked, a forced GALFIT
decomposition with bulge and disk ellipticities for our 1D 
model decomposition E to test whether the 1D and 2D solutions
differ only by virtue of their separate ellipticity and PA values.
The GALFIT parameters with variable (Model O) or forced (Model Q)
ellipticity certainly differ in the sense that the 2D (NLLS) Model Q
approaches the 1D (NLLS) Model F.  Some of the differences between
Models Q and, say, E, which can be as large as $\sim$ 10-20\%,
thus stem from using different component orientations (PAs) and
ellipticities.  Yet another reason to embrace 2D model decompositions.

Note that the total light fraction of the GALFIT bulge and disk
(B/D)-only fit is 29\% for the bulge and 71\% for the disk.
We also compute the total IRAC 3.6$\mu$m flux of M31 as
$M_{3.6}=-21.65$ or $L_{3.6}=9.8\times10^8 \lsol$. 

We can summarize the current state of model comparison as:
i) for simple data sets, the NLLS and MCMC methods yield
comparable results; ii) the MCMC method enables the
identification of data-model inadequacies; and iii), 
structural parameters can change by $\sim$ 10-20\% when
PA and ellipticity differences for the bulge and disk
viewed in 1D or 2D are taken into account. 
Finally, the differences between the 1D B/D decomposition parameters
in Table 2 for the 3.6$\mu$m IRAC and $I$-band Choi02 data are largely
due to waveband effects (see \se{results}).


\section{ANALYSIS OF THE COMPOSITE PROFILE}\label{sec:composite}

Having examined the validity of simple B/D model decompositions,
we now consider a more elaborate scheme with the decomposition
of our 1D $I$-band composite profile (\Fig{M31composite}) into
three structural components: a bulge, a disk and a halo.  From visual
inspection of \Fig{M31composite}, it is unclear which parameterization
best suits the faint data.  We only consider here a \sersic (\Eq{sersic})
or a power-law (\Eq{faint}) function.  The decomposition results are
reported in Table 4 for the power-law halo and in Table 5 for the
\sersic halo.

First, the comparison of Models R and S reminds us of MCMC's ability
to compensate for inadequate data-model matching.  
Contrary to Model R, MCMC's Model S calls for an extended, faint bulge
coupled with a very compact, bright disk.  The data-model
errors or bulge-disk model convariances are sufficiently large as to 
yield an uncertain decomposition.  The disk component is nearly
absent in this minor axis cut decomposition (reminiscent of PvdB94,
Irwin05's and G09's results).  The bulge component absorbs much of the
disk contribution, making the bulge completely dominant.  The
parameters of Model S are also significantly different than those
for Models R, T and U.  This situation may have led to the large
central $n$ values for the M31 bulge reported in the past.  Note
that the NLLS approach (Model R) forces a disk solution but the
fit parameters are never constrained. 

\subsection{To bin or not to bin?}
We can exploit different modeling techniques through a combination
of 1D NLLS and MCMC analysis with ``Binned'' or ``Unbinned'' cuts.
The ``unbinned'' method comes from our consideration of all the
data points in our four individual cuts (Choi02, Irwin02, PvdB94, G05)
treated individually.  The "binned" approach consists of averaging the
four data sets into one composite profile; the overlapping data
at any given radius are weight-averaged according to their
respective errors.  The differences in model parameters for
the ``Binned'' vs ``Unbinned'' data, as seen in the Model
pair S/T, can also be rather significant.
Unless a precise argument can be made for data binning
(e.g., improvements of the signal-to-noise ratio), the
latter is best avoided. 

\subsection{Power-law versus \sersic\ halo}

\Fig{M31composite_fits} show the composite profile decompositions
with Models U (power-law; left) and W (\sersic; right).  Both models
look statistically comparable given the similar residual distributions. 
Based on global $\chi$-square statistics for NLLS Models R and V,
which rely on minor axis data only, and visual impression of the
data-model residuals (figure not shown), the power-law fit seems
to be a marginally better description of the outer profile of M31.

However, the MCMC ``MinMaj'' Models U and W, which must also account
simultaneously for the major axis $I$-band data of Choi \etal (2002),
are intrinsically more rigorous than Models R and V.  The probability
distributions for the parameters of Model U and W are also very
similar.  And while the fit residuals in \Fig{M31composite_fits}
are essentially equivalent, we see that the disk-halo transition
for Model U occurs at $R_{min} \sim 8$ kpc while for Model W,
that transition is at $R_{\rm min} > 10 $ kpc.  This is made
even clearer when we plot, in \Fig{lightfraction}, the relative
light fraction of each component for Models U and W.  As we will
see in \se{fractions}, kinematic constraints for M31 red giants
(Gilbert \etal 2007; G09) suggest a sub-dominant disk for
$R_{\rm min} > 9 $ kpc and thus Model W could be disfavored
on those grounds alone.  Furthermore, with its high-$n$ index
for a \sersic\ halo, Model W calls for an excess of ``halo''
stars over disk stars in the inner parts of M31; the latter
has yet to be detected (e.g., Saglia \etal 2010) though the scales
involved are very small and both the measurements and their
interpretation would be very challenging.  For these reasons,
we favor the power-law description (Model U) for the halo component. 


\section{RESULTS}\label{sec:results}

We now analyse our results for the bulge, disk and halo
structural components.  The parameter values for the
various B/D decompositions of the IRAC data (Tables 2-3) are
presented graphically in \Fig{BDparams} as follows: 
MCMC (black), NLLS (red), and GALFIT (blue).
We also show the results for the NLLS B/D decompositions by
KB99 ($V$-band, grey annulus), Worthey etal (2005; $I$-band,
light green dot), and Seigar \etal (2008; 3.6$\mu$m, brown dot). 

The solutions for the Choi02 $I$-band profile decompositions
are not shown, for simplicity.  However, comparison of Models
I to N with Models A to G shows that the disk scale lengths,
$R_d$, are typically 20-25\% {\it larger} at shorter 
wavelengths\footnote{de Jong (1995) and M03 found similar
results for their larger collection of multiband luminosity
profiles of spiral galaxies.}.  Unfortunately, the wavelength dependence
of the bulge parameters is not as easily assessed with this limited
data base\footnote{The larger data base of M03 (see their Table 3)
yields no clear wavelength trends for the bulge index $n$ and 
$R_e$ either.}.  In any event, the analysis of both the stellar
populations and dust effects on scaling parameters is beyond the
scope of this analysis.  The main point to note here is that model
parameters extracted from our IRAC and $I$-band data may differ
due to wavelength effects.

\subsection{Bulge}\label{sec:bulge}

The bulge of M31, long-considered a prototypical de~Vaucouleurs
spheroid (e.g., Walterbos \& Kennicutt 1988; PvdB94; Irwin05)
is here confirmed to be less concentrated with a mean \sersic
shape index $n \simeq 2.2 \pm 0.3$ for all the models and data
reported in Tables 2-5.

Based on the IRAC luminosity profile at 3.6$\mu$m, the mean bulge
effective radius, $R_e$, projected along the major axis, is
$1.0 \pm 0.2$ kpc, and the mean bulge effective SB, $\mu_e$,
is $16 \pm 0.2$ \magarc.  The surface brightnesses are, again,
different for the $I$ and IRAC bands.

Given the known covariances between the parameters
$R_e$ and $\mu_e$ for spheroid systems (Mac03), it is
reassuring that the variations of three
bulge parameters for the different NLLS and MCMC codes
are so small.  For small bulges, the \sersic parameter
$n$ is not significantly covariant with other bulge
parameters (Fisher \& Drory 2010).  Indeed, we find that $n$
is independent of the decomposition method, data projection
(minor versus major cuts) but it does depend slightly on bandpass
($n$ being slightly larger at redder wavelength). 

Our results are consistent with the earlier report by KB99 that
$n = 2.19 \pm 0.10$ and $R_e = 1.0 \pm 0.2$ kpc for the
M31 bulge (shown as the grey annulus in \Fig{BDparams}). 
To our knowledge, KB99 is the first reported claim that
the M31 bulge is less cuspy than a de~Vaucouleurs profile
(see their \S{2}, p.777).  Their $V$-band decomposition
included a nucleus component but its impact on the bulge
parameters, as we saw, is negligible. 

Unaware of KB99, Worthey \etal (2005) decomposed the Choi02
$I$-band azimuthally-averaged surface brightness profile using
the Mac03 NLLS algorithm to find $n=2.0 \pm 0.2$ and $R_e = 1.0 \pm 0.3$ kpc
for the M31 bulge.  This solution matches our Model N.

Seigar \etal (2008) used the Spitzer/IRAC 3.6$\mu$m light
profile and a NLLS model decomposition to find for the bulge
$n=1.71 \pm 0.11$ and $R_e = 1.93 \pm 0.10$ kpc.  Their result
can be compared with our Model F; see also the brown dot in \Fig{BDparams}.
They find an extended, brighter, dominant bulge with $B/D=57\pm 2$\%, 
whereas ours is sub-dominant with $B/D \sim 43$\%.  We offer
no explanation for their derivation of an unusually large bulge
for M31. 

As seen in Table 2, the bulge ellipticities determined by MCMC,
with simultaneous major and minor axis analysis, and isophotal
contours average $\epsilon_b=0.25 \pm 0.4$ are clearly
at odds with $\epsilon_b=0.37$ fron GALFIT.  This discrepancy
is because, unlike our 1D profile analyses, GALFIT
accounts for PA variations, and the inner isophotal
contours are the sum of individual components (bulge
and disk) with different ellipticities and PAs.
The GALFIT PAs for the bulge and disk differ by 20\deg, 
whereas our MCMC code assumes co-aligned bulge and disk
components.  In Model Q, we see that forcing the MCMC ellipticities
into GALFIT yields parameters that are closer to Model F. 
One can then only assume that a Markov-Chain 2D decomposition
(to be tested elsewhere) would recover results similar to Model E. 
Finally, comparison of Tables 2 and 4 (Models L vs R and M vs U)
shows that inclusion of the halo component does not affect
the bulge light significantly (if at all, within the
decomposition errors).  

\subsection{Disk}\label{sec:disk}

The spread of disk solutions from model-to-model variations
might be expected to be larger than that of the bulge, in
light of the inherent sensitivity from code-to-code to
enhanced non-axisymmetric features.  However, uncertainties
in the disk measurements translate directly into bulge 
parameters.  Typically, published disk scale lengths differ
by $\sim 20$\% from author-to-author (Mac03).  \Fig{BDparams}
however shows that our disk scale length solutions differ 
by less than that (at the one-sigma level). 

Our basic exponential disk model applied to the IRAC data
has a scale length, $R_d$, of $5.3 \pm 0.5$ kpc.  The rather
small 10\% uncertainty reflects the range of our model
solutions from MCMC to 1D NLLS and 2D NLLS (GALFIT), 
i.e., from short to longer values.  Proper accounting
of model-data discrepancies, as with MCMC, yields shorter
disk scale lengths.  A realistic disk scale length
for M31, measured in a band dominated by the older, low
mass disk stars, probably lies closer to 5 kpc.

For their B/D decomposition of $V$-band luminosity profiles,
KB99 obtained $R_d = 1698$\arcsec = 6.5 kpc.  
We know from Mac03 that $V$-band disk scale lengths
are roughly 10\% longer than $I$-band and 25\% longer
than $H$-band disk scale lengths.  Thus, barring any additional
colour effect from $H$ and 3.6$\mu$m, the 6.5 kpc $V$-band
scale length is equivalent to 5.2 kpc at 3.6$\mu$m and
5.9 kpc at $I$-band in very good agreement with Models F and N.
Worthey \etal (2005) measured an $I$-band value of 
$R_d = 5.8 \pm 0.2$ kpc (adjusted to a distance of 785 kpc), 
identical to Model N.  Seigar \etal (2008) obtained
$R_d = 5.91 \pm 0.27$ for the IRAC 3.6 $\mu$m profile
which compares well with our Model F. 

In the spirit of McD09, we also explored
the effect of spiral arm clipping to recover the
underlying old stellar population disk, as discussed
in \se{methods}. 
Such clippings typically yield a shallower scale length,
at least for 1D SB profiles (\se{methods}).  

Given the very large number of disk pixels, and negligible 
bulge light into the disk beyond $R_{min} \simeq 1.2$~kpc,
the disk ellipticity is well constrained by our different
analyses (see Tables 2 \& 3).  
Whether through isophotal fits, MCMC or GALFIT, we find
for the IRAC data $\epsilon_d = 0.73 \pm 0.01$.  This
ellipticity corresponds to a dust-free, razor-thin disk
projected at a 74\deg angle.  Assuming a thickness for the M31
disk corresponding to $c/a=.175$ (Haynes \& Giovanelli 1984), 
the M31 viewing angle is then 77.6\deg.  

As we shall see in \Fig{lightfraction}, the disk extends out
to roughly $R_{min} \sim 15$~kpc or $R_{maj} \simeq 56$~kpc.
Other lines of evidence (e.g., Ibata \etal 2005; Worthey \etal 2005)
point to the M31 stellar disk extending out to $R_{maj} \sim 50$ kpc,
with occasional detections of features (not spatially continuous)
with disk-like kinematics out to $R_{maj} \sim 60$ kpc (Ibata \etal 2005).  
The latter is far beyond the disk-halo transition at $R_{\rm min} \sim 8$~kpc
(see \Fig{lightfraction}). 

\subsection{Halo}\label{sec:halo}

The deep star counts accentuate the presence of the halo 
at $R_{\rm min} > 10$ kpc that is seemingly neither the extension
of the inner low-$n$ bulge or the exponential disk, nor the
manifestation of a faint de~Vaucouleurs halo (Tanaka \etal 2010
fit their extended Subaru Suprime-Cam surface brightnesses of the
M31 stellar halo with a Hernquist profile and scale radius equal
to 14 kpc).  As we saw in \se{composite}, the \sersic model W was
rejected on account of kinematic constraints.  Model U, shown in
\Fig{M31composite_fits}, appears to be a better representation
of the halo star counts.  Note that the disk and faint components
are equal at R$_{\rm min} \sim 8-9$ kpc. 

The star counts are consistent with a halo structure best
described with a 2D spatial density distribution power-law index 
$-2.5 \pm .2$ ($-2\times\alpha$ as defined in \Eq{faint}). 
In 3D, the halo can thus be approximated as a sphere with
$\log\rho(r) = {-3.5}\log{r}$.
The solution from Model U agrees well with other
power-law models of the M31 stellar halo (Irwin05; Guha05;
Ibata \etal 2007; G09; Tanaka \etal 2010).  Tanaka \etal (2010)
obtained deep Subaru imaging of M31 along both minor axes, north
and south, to find a 2D power-law index equal to $-2.17\pm0.15$,
in excellent agreement with our value.  Our combined results are
marginally consistent with Ibata \etal (2007) who measured $-1.91\pm 0.11$
for their halo power-law fit.  

A comparison with the Milky Way is worthwhile. For instance,
Kenner \etal (2008) modeled the number density of RR Lyrae stars
along the ecliptic over the range $10<R<45$~kpc as a power law with
index $n=-2.48 \pm 0.09$.  Carollo \etal (2010) also identified
inner- and outer-halo components of the Galaxy based on star counts,
kinematics and abundances of 32360 SDSS and SDSS-II stars.
The transition between each component would be around 15 kpc (along
the minor axis).  The slopes of the power-law spatial density distributions
of their inner- and outer-halo are $-3.17 \pm 0.20$ and $-1.79 \pm 0.29$,
respectively.  The density profile for their measured outer halo
is substantially shallower than that of their inner halo on account
of the higher velocity dispersion and higher flattening of the former.   
Overall, the MW outer-halo stars possess a wide range of orbital
eccentricities, exhibit a clear retrograde net rotation, and are
drawn from a metallicity distribution function that peaks at
[Fe/H] = -2.2 (Carollo \etal 2010). 

Considering its transition range and power-law slope, our
halo component for M31 seems analogous to that of the Milky Way
outer halo component identified by Carollo, Kenner and others. 
Gilbert \etal (2007; 2009b) find the velocity dispersion of M31
halo stars in the range 30-60 kpc in excess of 100 \kms.  This 
is not only consistent with halo kinematics, it also matches 
the values quoted by Carollo \etal (2010) for the MW outer halo. 

Comparisons between the Milky Way and M31 are still challenging
considering the few available lines of sight for the measurements
of M31 kinematics (Gilbert \etal 2007; 2009).  As a result, testing
for the rotation of M31's halo is still beyond reach.  However, 
the M31 halo appears to be more metal-rich
($\langle [{\rm Fe/H}]\rangle=-0.47\pm 0.03$)
closer to its center (around $R_{\rm min}=11-20$ kpc), and more 
metal-poor further out ($\langle {\rm Fe/H} \rangle =-0.94\pm 0.06$ at 
$R_{\rm min}\sim 30$ kpc, and $\langle {\rm Fe/H} \rangle =-1.26\pm 0.1$
at $R_{\rm min}>60$ kpc) (e.g., Kalirai \etal 2006; Koch \etal 2008).  
This qualitatively resembles the findings from Carollo \etal (2010)
that the Milky Way's ``inner halo'' is more metal-rich than its
``outer halo''.  Their ``inner halo'' ($R_{\rm min}=10-15$ kpc) has
a metallicity distribution that peaks at  $\langle {\rm Fe/H} \rangle =-1.6$,
while that of their ``outer halo'' ($R_{\rm min}=15-20$ kpc) peaks at
$\langle {\rm Fe/H} \rangle = -2.2$.  Note that the stars in their
Milky Way ``outer halo'' component are within the radial range of
the innermost M31 fields of Gilbert \etal (2007; 2009). 
The metallicity of M31's halo at large radii is still quite uncertain:
whether it becomes increasingly metal-poor with radius, or if it
continually decreases out to, say, $R_{\rm min}\sim50$ kpc and then
levels off is still a point of contention.

Numerous studies of the Milky Way halo based on the distribution 
and kinematics of RR Lyrae, blue horizontal branch, and giant
halo stars over different radial ranges report a fraction of
stellar halo light to total galaxy light of $\sim 1$\% (e.g.,
Morrison 1993; Chiba \& Beers 2000; Purcell \etal 2007; Bell \etal 2008).
This number grows to 2\% if unbound Sagittarius stream stars
are included (Law \etal 2005).  Considerable Milky Way halo
light thus comes from dynamically young material in various
tidal streams.  By contrast, our power-law models for the M31
halo (Table 4, Models R-U and \Eq{totalhalo})
yield a fraction of halo-to-total light of $\ltsim 4$\%.  
This value is an upper bound as we assume the continuation
of the halo power-law density well into the bulge and disk
of M31.  The total light of the halo is integrated out 
to 200 kpc.  

Purcell \etal (2008) studied the accretion histories of
systems ranging from small spiral galaxies to rich galaxy
clusters.  \Fig{HaloLight} shows the predicted light 
fraction of the diffuse halo stars against the total
galaxy light for a variety of accretion histories which
account for much of the inner halo light, as a function
of host halo mass.  Overplotted with blue and red arrows
are the measured light fractions, $\sim 2\%$ and $\sim 4\%$
for the Milky Way and M31 respectively.  The masses for
the Milky Way ($\ltsim 10^{12} \msol$) and M31
($\sim 1.3 \times 10^{12} \msol$) are taken from
Courteau and van den Bergh (1999)\footnote{For a more
recent, but compatible, estimate of the MW mass, see
Busha \etal (2010).}. 
Our results are thus consistent with a common evolution
of the old stellar systems in the Galaxy and the Milky Way:
the inner regions ($R < 20-25$ kpc) would contain both
accreted and in situ stellar populations while the
outer regions ($R > 25-30$ kpc) would be assembled
through pure accretion and the disruption of satellites
(Carollo \etal 2007; Purcell \etal 2008; Carollo \etal 2010
and references therein).

\subsection{Component Total Light Fractions}\label{sec:fractions}

Using Eqs. 3, 5, 7, and 8, we can compute the light fractions
of the \sersic bulge, exponential disk, and power-law halo at
any radius as shown in \Fig{lightfraction} for Models U and W.
We also first compute the nucleus total light fraction using
KB99's exponential model (\se{nucleus}).  We find that the 
nucleus contribute a mere 0.05\% of the total galaxy light out
to 22 kpc (extent of the IRAC 3.6$\mu$m 1D profile along the
major axis).  We will ignore the nucleus component, and its
negligible contribution to the total luminosity profile, in 
measurements below. 

Using the IRAC 3.6$\mu$m 1D brightness profiles, we find that
the bulge-and-disk only fit yields $\sim 21$\% of the bulge
light and 79\% of the disk light, for a B/D ratio of 0.27 out
to $R\simeq 22$ kpc. 
Our 2D decompositions of the IRAC image (see \se{methods})
yield light fractions for the bulge and disk equal to
29\% and 71\% respectively, for a B/D ratio of 0.41. 

A halo fraction can be estimated from the decomposition of 
the composite profiles; see Tables 4 and 5. 
Using Model U, we find that the bulge, disk and halo each
contribute roughly 23\%, 73\% and 4\% of the total light
of the galaxy out to 200 kpc along the minor axis.
These light fractions are roughly similar to those obtained
with B/D-only fits within 23 kpc along the major radius; 
this is because the amount of additional bulge and disk light
beyond that radius is small and the halo profile takes over
at that point.

\Fig{M31composite_fits} and \Fig{lightfraction} show
that the light fraction of the galaxy components are consistent
with a dominant bulge within R$_{\rm min} \ltsim 1$ kpc, a dominant
disk between $1.5 \ltsim$ R$_{\rm min} \ltsim 8$ kpc, and a
dominant halo beyond R$_{\rm min} \gtsim 9$ kpc.
Note that the latter result is marginally consistent with the
light fractions derived by Kalirai \etal (2006), Gilbert \etal (2007), 
and G09 based on their measurement of stellar kinematics
along the minor axis of the M31 halo.  Their results show that
a disk is undetected as a cold component on the minor axis at
$R_{\rm min}\sim 9$ kpc.
Taken at face value, the Gilbert \etal constraint implies
that much of the light at 9 kpc ascribed to the disk in our
decompositions must come from either the bulge or faint components.  

Subtle differences between our photometric and spectroscopic analyses
could be reconciled if various factors are accounted for.  For one,
our photometric and spectroscopic data were extracted in different
ways and may weigh the stellar density field differently.  
For instance, our light profile wedges do not sample the same lines
of sight as those for the spectroscopically mapped star counts.

The photometrically-(this Paper) and kinematically-based decompositions
may also be sensitive to the M31 structure in different ways.  The disk
structure at $R_{min} \simeq 9$~kpc (roughly 35~kpc along the major axis)
in the disk plane may be relevant.  The disk inclination and position
angle are assumed to be constant to these large radii and the kinematic
analysis assumes a cold disk.  However, at 5-6 disk scale lengths, disks
can warp, flare, etc.  Resolution of any discrepancies between our
respective decompositions ultimately relies on a {\it simultaneous}
analysis of the photometric and spectroscopic data.


\section{DISCUSSION and CONCLUSIONS}\label{sec:discussion}

\subsection{Luminosity Decomposition Methods}

Our paper has provided an investigation of luminosity profile 
decomposition methods.  Whether surface brightness data are
(i) binned or not, (ii) treated in magnitudes or counts, 
(iii) clipped for non-axisymmetric features, (iv) analysed
in 1D or 2D, (v) originate from wedges (cuts) or azimuthally-averaged
profiles, and/or (vi) have different radial ranges or are modeled
using different minimization schemes (e.g., MCMC vs NLLS), may
yield slightly different structural parameters.  The uniqueness
of model parameters in bulge+disk+halo decompositions
is also often thwarted by parameter covariances.  Despite these
caveats, we have achieved self-consistent decomposition results.
The technical considerations about light
decomposition methods presented in this paper (see also Mac03)
suggest systematic errors of galaxy structural parameters from
method-to-method (and author-to-author) of order 20\%.

Perhaps the most significant difference between 1D and 2D model
parameters (fitted over the same radial range) comes from the ability
to decompose the position angles and ellipticities of two 
photometrically-distinct superimposed systems.  This crucial
operation is only possible in 2D.  Our simultaneous fitting
of the minor and major axes in 1D has also provided independent
information about bulge and disk ellipticities, but not position angle.
Future investigations of the M31 2D light distribution will
require extensive MCMC decompositions which are currently
computationally expensive (e.g., Yoon \etal 2010). 

\subsection{Deep Light Profiles and Galaxy Halos}

The extended M31 galaxy light profile with deep star counts 
reveals a halo component beyond $R_{\rm min} \simeq 10$ kpc at a brightness 
level $\mu_I \sim 27$ \magarc\ (\Fig{M31composite}; also Guha05, 
Ibata \etal 2007; Tanaka \etal 2010). 
Because new, deep galaxy surveys\footnote{Such as the Next Generation
Virgo Cluster Survey (Ferraresse et~al., in prep.) or other galaxy surveys
by PanSTARRS or with the Large Synoptic Survey Telescope, to name a few.}
should reach at least a few orders of magnitude below this brightness
threshold, the common detection of galaxy halos should soon be within reach.   
If M31 is at all representative, this is important for galaxy structure
studies since the decomposition of deep galaxy light profiles will likely
be skewed if the halo component is ignored.

M31's $n=2.2$ bulge is mostly dominant over the range $R_{\rm min} < 1.2$ kpc.
The disk takes over in the range 1.2 kpc $< R_{\rm min} < 9$ kpc, whereas
the halo likely dominates the light for $R_{\rm min}> 9 $ kpc
(see \Fig{M31composite_fits}).
Our intimation of disk dominance between 1.2 and 9 kpc on the minor axis
is only marginally consistent with stellar kinematics of the M31 halo
whereby the disk is completely cold and absent at 9 kpc (Gilbert \etal 2007;
Gilbert09a).  
We stress that our decompositions are based on photometric information
alone; additional information about the overall structure of M31, such
as radial abundance variations or the complete kinematic mapping of M31's
major structural components, may possibly 
yield somewhat different decompositions than the ones presented here.
Any tension that may exist between the structural decompositions derived
from our photometric analysis and separate kinematic analyses can only
be resolved with a simultaneous modeling of the two data sets.  Such an
exercise is beyond the scope of the present paper but will undoubtedly
drive future investigations of the M31 galaxy structure.

\subsection{Bulge Formation}

The ratio of bulge-to-disk scale lengths, $R_e/R_d$, 
has been viewed as a telltale signature of galaxy secular
evolution (Courteau, de Jong, \& Broeils 1996; Mac03; 
Kormendy \& Kennicutt 2004).  Mac03 showed that most
nearby spiral galaxies have a B/D ratio $R_e/R_d\simeq 0.2$
which is a natural prediction of secular evolution models
(Courteau, de Jong, \& Broeils 1996).  It is thus remarkable
that, for M31, $R_e/R_d\simeq 0.2$.  Furthermore, a value
$n < 2$, as for M31, is often interpreted as evidence
for secular evolution (Kormendy 1993; Courteau \etal 1996;
Fisher \& Drory 2010).  M31's $R_e/R_d$ ratio and bulge
\sersic index close to 2 suggests some amount of secular
evolution for the bulge. 

However, a discussion of the formation scenarios of the M31
bulge cannot simply rely on the shape of its luminosity profile.
For instance, the latter may be biased by a recent frosting of star
formation (e.g., de Jong \& Lacey 2000; MacArthur etal 2004).  The
detailed investigation of bulge formation scenarios requires
resolved spectroscopy.  MacArthur etal (2009) used deep Gemini/GMOS
spectra to show that numerous late-type spirals with nearly-exponential
($n\sim1$) bulges harbor mostly old stellar populations.  This is
confirmed for M31 by Saglia etal (2010) whose long-slit spectra
centered on the M31 bulge show a predominantly old ($\geq$ 12 Gyr),
slightly $\alpha$-element enhanced ($[\alpha/\rm{Fe}] \approx +0.2$), 
solar metallicity bulge.
In the inner arcseconds, the M31 luminosity-weighted age
drops to 4-8 Gyr and the metallicity increases to three times
solar. A possible formation scenario thus involves a very
rapid, early, assembly of the M31 bulge followed by subsequent
accretion via secular transport of enriched, disk material
to the center.  This is neither a purely classical or 
secular evolution formation scenario, but rather one 
where both mechanisms are at play.  Similar conclusions have
been drawn for the Milky Way; Babusiaux \etal (2010) report
the presence of two distinct populations in Baade's Window
with a metal-rich population with bar-like kinematics and
a metal-poor population with kinematics indicative of an old
population.  The entangled interpretation of spiral galaxy bulges,
like M31's, in terms of both merger-induced (classical) and secular
evolution for larger samples of galaxies is also discussed in
Peletier \etal (2007), MacArthur \etal (2009) and Roediger \etal
(2011), among others. 

Numerous questions remain about the spatial distribution,
populations and kinematics of M31 stars.  However, one conclusion
is clear: M31 likely evolved through a rich past that involved early
``classical'' merging and structural readjustements via secular
evolution effects and satellite accretion.  As a result, M31's
luminosity profile does not follow a simple de~Vaucouleurs law.

\bigskip We wish to thank Pauline Barmby and Mike Irwin for providing
electronic copies of the Spitzer IRAC images from Barmby \etal (2006)
and minor axis deep counts from Irwin \etal (2005), respectively.
We extracted our own profile cuts and azimuthally-averaged profiles
from the IRAC images as presented in Figures 1, 3, and 4.  Chris Purcell
and James Bullock also very kindly contributed \Fig{HaloLight}.
We are grateful to John Kormendy for sharing unpublished model
parameters about the Kormendy \& Bender (1999, KB99) analysis
and for additional comments.  Final thanks go to Jonathan Sick
for computing the 2MASS~6X image zero points, Chien Peng for
thoughts about GALFIT parameter errors, and Daniela Carollo
and Tod Lauer for comments about their respective papers.  
S.C. and L.M.W. acknowledge support through respective Discovery
grants from the Natural Sciences and Engineering Research Council
of Canada.  P.~G. is also grateful to the NSF for support through
grants AST-0606932 and AST-1010039.  Lastly, K.M.G. is supported
by NASA through a Hubble Fellowship grant HST-HF-51273.01 awarded
by the Space Telescope Science Institute, which is operated by
the Association of Universities for Research in Astronomy, Inc.,
for NASA, under contract NAS 5-26555.


\clearpage


\begin{deluxetable}{lccc} 
\tabletypesize{\scriptsize}  
\tablewidth{0pc}  
\tablecaption{Sources of surface brightness profiles for M31} 
\tablehead{
\colhead{Reference} & 
\colhead{Bands} & 
\colhead{pixel scale} & 
\colhead{min. exp. time}}
\startdata 
Walterbos \& Kennicutt (1987)    & $UBVR$   & photographic plate   & 300-7200s \\
Pritchet \& van den Bergh (1994) & $V$      & star counts          & 2700s \\
Choi \etal (2002)                & I        & 2.03 $\arcsec$/pixel & 2400s \\
Irwin \etal (2005)               & Gunn $i$ & star counts          & 800-1000s \\
Beaton \etal (2007)  2MASS 6X    & JHK      & 1.00 $\arcsec$/pixel & 46.8 sec \\
Barmby \etal (2007)  IRAC        & 3.6$\mu$m, 4.5$\mu$m & 0.863 $\arcsec$/pixel & 62-107s \\
Gilbert \etal (2009)             & V        & spectroscopic star counts & 3600 s \\
\enddata 
\label{tab:tab1}
\end{deluxetable} 

\bigskip

\begin{deluxetable}{lcllccccccc}
\tabletypesize{\scriptsize} 
\tablewidth{0pc}  
\tablecaption{Bulge/Disk Decompositions for M31}
\tablehead{
\colhead{} &
\colhead{Data} &
\colhead{Method} &
\colhead{Cut} &
\colhead{$n$} &
\colhead{$R_e$} &
\colhead{$\mu_e$} &
\colhead{$\epsilon_{\rm bulge}$} &
\colhead{$R_d$} &
\colhead{$\mu_0$} &
\colhead{$\epsilon_{\rm disk}$} \cr
\colhead{} &
\colhead{} &
\colhead{} &
\colhead{} &
\colhead{} &
\colhead{(kpc)} &
\colhead{(\magarc)} &
\colhead{} &
\colhead{(kpc)} &
\colhead{(\magarc)} &
\colhead{} \\
}
\startdata
A. &IRAC &MCMC &Maj & $2.46 \pm 0.09$ & $1.09\pm 0.07$ & $16.2 \pm 0.10$ & $--$ & $5.09\pm 0.17$ & $16.83\pm 0.07$ & $--$  \cr 
B. &IRAC &NLLS &Maj & $2.3  \pm 0.3 $ & $1.0 \pm 0.2 $ & $16.1 \pm 0.3 $ & $--$ & $5.4 \pm 0.3 $ & $16.8\pm 0.1$   & $--$  \cr 
C. &IRAC &MCMC &Min & $2.01 \pm 0.11$ & $0.53\pm 0.05$ & $15.47\pm 0.13$ & $--$ & $1.29\pm 0.09$ & $16.49\pm 0.25$ & $--$  \cr
D. &IRAC &NLLS &Min & $1.9 \pm 0.5$  & $0.5 \pm 0.3 $ & $15.4\pm 0.5 $  & $--$ & $1.3\pm 0.6  $ & $16.4\pm .15 $  & $--$  \cr
E. &IRAC &MCMC &MinMaj & $2.18\pm 0.06$& $0.82\pm0.04$ & $15.77\pm 0.07$ & $0.21\pm 0.01$ & $4.71\pm 0.14$ & $16.62\pm 0.05$ & $0.74\pm 0.01$ \cr
F. &IRAC &NLLS &AZAV & $2.4\pm 0.2$ & $1.10\pm 0.10$ & $16.1\pm 0.10$ & $--$ & $5.8\pm 0.1$ & $16.79\pm 0.02$ & $0.74\pm 0.02$ \cr
G. &IRAC &MCMC &AZAV & $1.66\pm 0.03$ & $0.68\pm 0.01$ & $15.34\pm 0.03$ & $--$ & $4.75 \pm 0.01$ & $16.41\pm 0.01$ & $-- $ \cr
H. &IRAC &NLLS &AZAVmsk & $2.2\pm 0.3$ & $1.00\pm 0.10$ & $16.0\pm 0.20$ & $--$ & $4.9\pm 0.1$ & $16.70\pm 0.10$ & $0.74\pm 0.02$ \cr
       &         &      &               &            &  &  &  &  \cr 
I. &Choi02 &MCMC &Maj & $2.06\pm 0.06$  & $0.91 \pm 0.04$ & $18.12\pm 0.08$ & $--$ & $5.69\pm 0.09$ & $18.97 \pm 0.03$ & $--$  \cr 
J. &Choi02 &NLLS &Maj & $2.2 \pm 0.3$   & $1.12 \pm 0.10$ & $17.3 \pm 0.2$  & $--$ & $6.4 \pm 0.1$  & $18.14 \pm 0.04$ & $--$  \cr 
K. &Choi02 &MCMC &Min & $1.85\pm 0.07$  & $0.51 \pm 0.02$ & $17.62\pm 0.08$ & $--$ & $1.73\pm 0.05$ & $18.99 \pm 0.08$ & $--$  \cr 
L. &Choi02 &NLLS &Min & $1.9 \pm 0.2$   & $0.53 \pm 0.03$ & $17.7 \pm 0.1$  & $--$ & $1.68\pm 0.04$ & $18.8  \pm 0.1$  & $--$  \cr 
M. &Choi02 &MCMC &MinMaj & $1.83\pm0.04$& $0.74 \pm 0.02$ & $17.73\pm 0.05$ & $0.28\pm 0.01$ & $5.47 \pm 0.08$ & $18.91\pm 0.03$ & $0.70\pm0.01$  \cr
N. &Choi02 &NLLS &AZAV & $2.00 \pm 0.4$ & $1.00 \pm 0.30$ & $18.2 \pm 0.4$  & $--$ & $5.80 \pm 0.10$ & $19.0 \pm 0.3$ & $0.71\pm 0.02$ \cr 
\enddata
\label{tab:tab2}
\end{deluxetable}

\bigskip 

\begin{deluxetable}{llcccccccc}
\tabletypesize{\scriptsize}  
\tablewidth{0pc}  
\tablecaption{GALFIT Bulge/Disk Decompositions of the M31 IRAC 3.6 $\mu$ image 
\break (parameter errors are typically 10\%)} 
\tablehead{
\colhead{} &
\colhead{Method} &
\colhead{$n$} &
\colhead{$R_e$} &
\colhead{$\mu_e$} &
\colhead{$\epsilon_{\rm bulge}$} &
\colhead{$R_d$} &
\colhead{$\mu_0$} &
\colhead{$\epsilon_{\rm disk}$} \cr
\colhead{} &
\colhead{} &
\colhead{} &
\colhead{(kpc)} &
\colhead{(\magarc)} &
\colhead{} &
\colhead{(kpc)} &
\colhead{(\magarc)} &
\colhead{} \\
}
\startdata
O. &Original Image & $1.9$ & $0.9$ & $16.1$ & $0.37$ & $5.9$ & $17.1$ & $0.72$ \cr
P. &Masked   Image & $2.0$ & $1.0$ & $16.2$ & $0.37$ & $5.9$ & $17.1$ & $0.72$ \cr
Q. &Forced $\epsilon_{\rm bulge}$, $\epsilon_{\rm disk}$ & $2.1$ & $0.9$ & $16.9$
   & $0.21$ & $5.7$ & $16.8$ & $0.74$ \cr
\enddata
\label{tab:tab3}
\end{deluxetable}


\bigskip


\begin{deluxetable}{lllcccccccc}
\tablecaption{$I$-band composite; Bulge/Disk/Power-law Faint} 
\tabletypesize{\scriptsize}  
\tablewidth{0pc}  
\tablehead{ 
\colhead{} & 
\colhead{Method} &  
\colhead{Cut} &  
\colhead{$n$} & 
\colhead{$R_{e,b}$} &  
\colhead{$\mu_{e,b}$} &  
\colhead{$R_d$} & 
\colhead{$\mu_0$} & 
\colhead{$\alpha$} & 
\colhead{$\mu_*$} & 
\colhead{$a_h$} \cr 
\colhead{} & 
\colhead{} & 
\colhead{} & 
\colhead{} & 
\colhead{(kpc)} & 
\colhead{(\magarc)} & 
\colhead{(kpc)} & 
\colhead{(\magarc)} & 
\colhead{} & 
\colhead{(\magarc)} & 
\colhead{(kpc)} \\
}  
\startdata 
R. &NLLS &Min/Bin      & $2.20^{+.20}_{-.20}$ & $0.60^{+.10}_{-.10}$ & $17.90^{+.30}_{-.30}$ & $1.70^{+.30}_{-.30}$ & $18.80^{+.30}_{-.30}$ & $1.10^{+.10}_{-.10}$ & $27.70^{+.10}_{-.10}$ & $11.40^{+0.10}_{-0.10}$ \\ 
S. &MCMC &Min/Bin     & $2.67^{+.11}_{-.11}$ & $1.77^{+.05}_{-.05}$ & $19.53^{+.07}_{-.07}$ & $0.12^{+.014}_{-.014}$& $15.60^{+.20}_{-.20}$                 & $1.78^{+.29}_{-.30}$ & $27.90^{+.08}_{-.08}$ & $55.65^{+12.3}_{-12.6}$ \\ 
T. &MCMC &Min/UnBin   & $1.89^{+.10}_{-.10}$ & $0.51^{+.03}_{-.03}$ & $17.66^{+.11}_{-.11}$ & $1.54^{+.05}_{-.05}$ & $18.72^{+.09}_{-.09}$ & $1.14^{+.09}_{-.09}$ & $28.32^{+.09}_{-.09}$ & $6.60^{+1.22}_{-1.20}$ \\ 
U. &MCMC &MinMaj/UnBin   & $1.90^{+.05}_{-.05}$ & $0.73^{+.02}_{-.02}$ & $17.79^{+.05}_{-.05}$ & $5.02^{+.05}_{-.06}$ & $18.81^{+.02}_{-.02}$ & $1.26^{+.04}_{-.04}$ & $28.07^{+.05}_{-.05}$ & $5.20^{+0.16}_{-0.16}$ \\ 
\enddata 
\label{tab:tab4}
\end{deluxetable} 


\bigskip 



\begin{deluxetable}{lllcccccccc}
\tabletypesize{\scriptsize}  
\tablewidth{0pc}  
\tablecaption{$I$-band composite; Bulge/Disk/\sersic Faint}
\tablehead{ 
\colhead{} &
\colhead{Method} &  
\colhead{Cut} &  
\colhead{$n$} & 
\colhead{$R_{e,b}$} &  
\colhead{$\mu_{e,b}$} &  
\colhead{$R_d$} & 
\colhead{$\mu_0$} & 
\colhead{$n_f$} & 
\colhead{$\mu_{e,f}$} & 
\colhead{$R_{e,f}$}  \cr 
\colhead{} & 
\colhead{} & 
\colhead{} & 
\colhead{} & 
\colhead{(kpc)} & 
\colhead{(\magarc)} & 
\colhead{(kpc)} & 
\colhead{(\magarc)} & 
\colhead{} & 
\colhead{(\magarc)} & 
\colhead{(kpc)} \\
}  
\startdata 
V. &NLLS &Min/Bin & $2.00^{+.30}_{-.30}$ & $0.50^{+.10}_{-.10}$ & $17.70^{+.20}_{-.20}$ & $1.40^{+.20}_{-.20}$ & $18.60^{+.30}_{-.30}$ & $4.90^{+0.80}_{-0.80}$ & $24.70^{+2.0}_{-2.0}$ & $9.00^{+1.3}_{-1.3}$ \\ 
W. &MCMC &MinMaj/Unbin & $1.80^{+.05}_{-.05}$ & $0.69^{+.02}_{-.02}$  & $17.72^{+.05}_{-.05}$ & $5.05^{+.05}_{-.06}$ & $18.80^{+.02}_{-.02}$ & $5.39^{+0.79}_{-0.80}$ & $26.11^{+.08}_{-.08}$ & $13.36^{+0.5}_{-0.5}$ \\ 

\enddata
\label{tab:tab5}
\end{deluxetable} 


\newpage


\clearpage

\clearpage

\clearpage

\clearpage

\newpage

\newpage
\begin{figure*}
\begin{center}
\begin{tabular}{cc}
\includegraphics[width=0.5\textwidth]{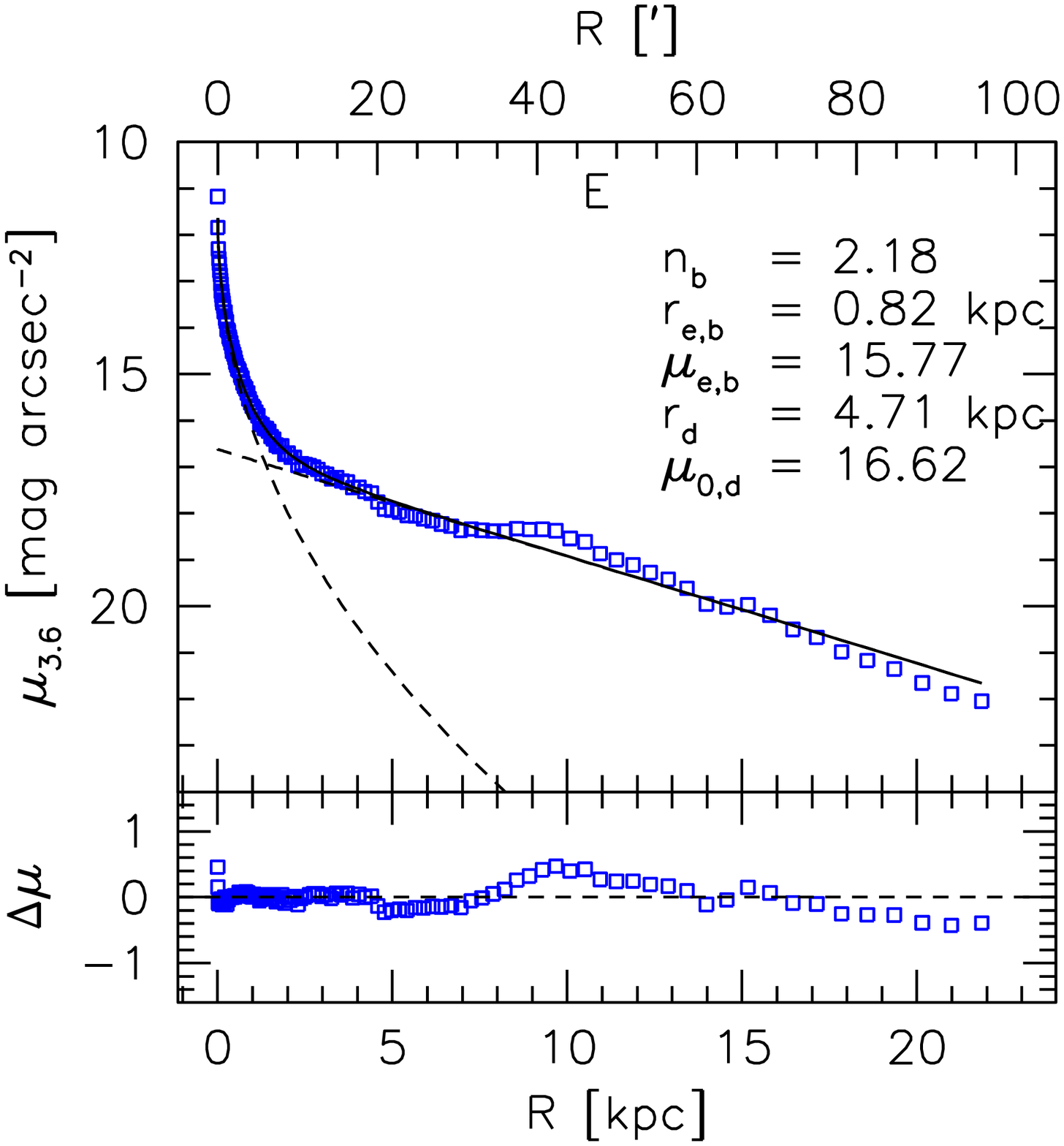} & \includegraphics[width=0.5\textwidth]{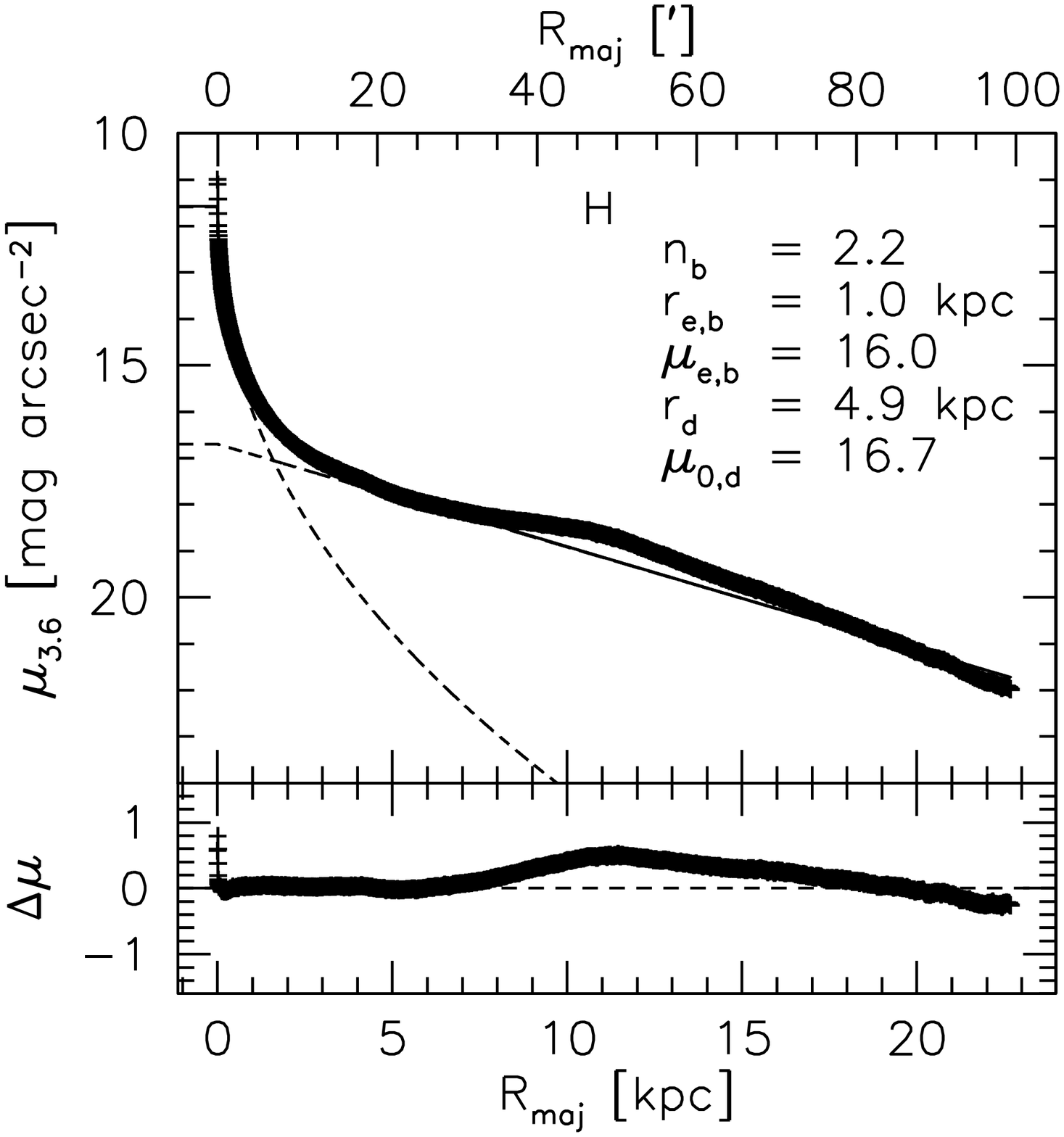}\\
\end{tabular}
\caption{(Left) MCMC decomposition of the IRAC 3.6$\mu$m minor and major
axis cuts for M31 with a \sersic\ bulge and an exponential disk model.  
The blue squares show the major-axis cut and the dashed lines are the
fitted bulge and disk component of Model E in Table 2.  The spike near
the center ($R\sim0$) in the data-model residuals (bottom panel) is due
to the nucleus. (Right) NLLS decomposition of the IRAC 3.6$\mu$m 
azimuthally-averaged surface brightness profile (with masked arms),
here shown with black crosses, with a \sersic\ bulge and an exponential
disk model from Model H in Table 2.  The data-model residuals are also
shown in the bottom panel.  Despite using two different decomposition
methods, Models E and H are virtually identical (within the errors).
}
\label{fig:M31_BDdecomp}
\end{center}
\end{figure*}

\newpage
\begin{figure*}
\begin{center}
\begin{tabular}{cc}
\includegraphics[width=1.0\textwidth]{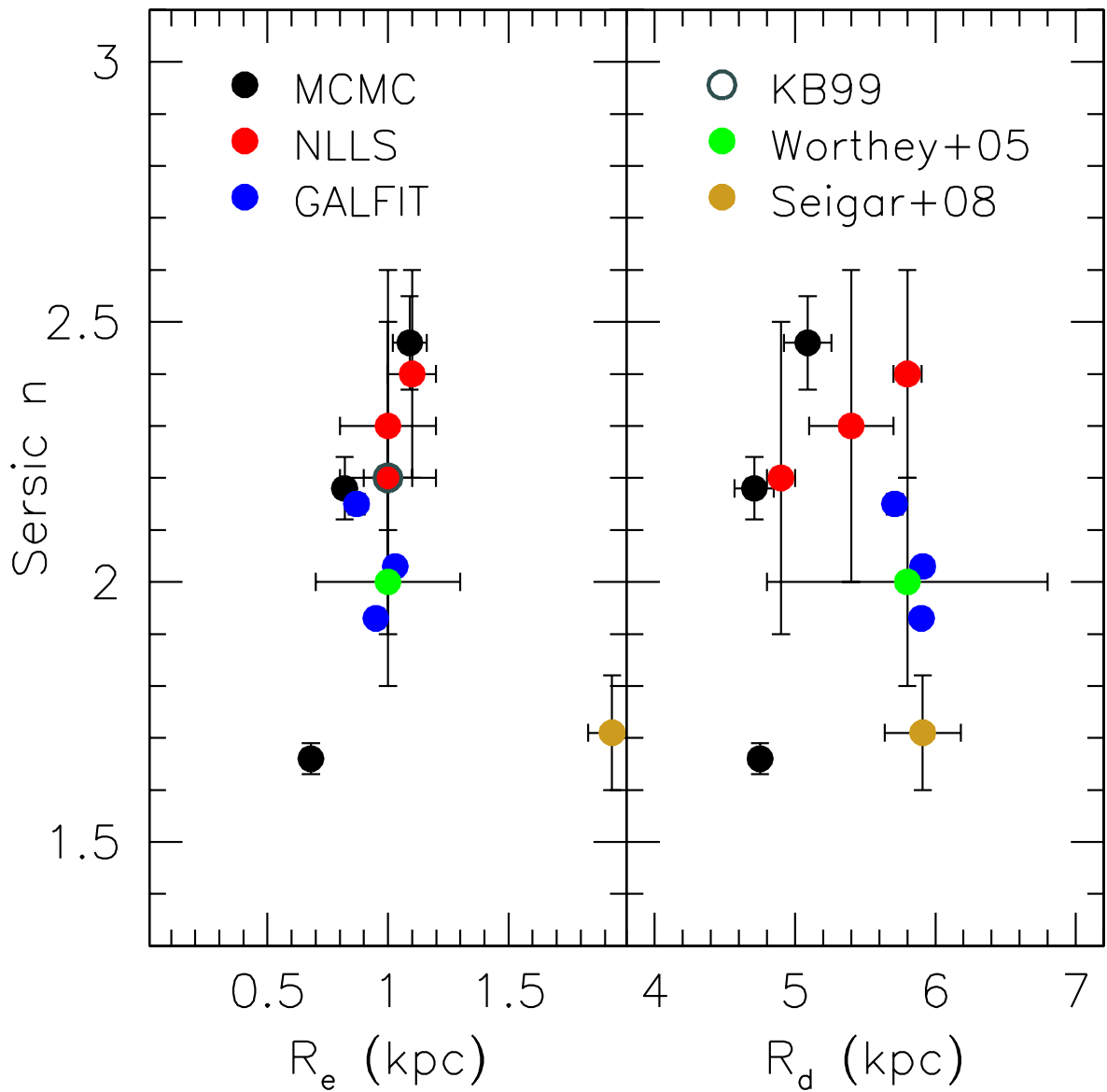}
\end{tabular}
\caption{Distribution of model parameters for the
B/D decompositions of the IRAC data as listed in Tables 2 \& 3.
Black points are for the MCMS Models A, E, and G.  Red points
are for the NLLS Models B, F, and H.  Blue points are for
the GALFIT Models O, P, and Q.  The grey annulus, green and
brown dots are the Kormendy \& Bender (1999), Worthey \etal (2005),
and Seigar \etal (2008) solutions, respectively.  The X-axes 
correspond to the face-on projection.}
\label{fig:BDparams}
\end{center}
\end{figure*}

\newpage
\begin{figure*}
\begin{center}
\begin{tabular}{cc}
\includegraphics[width=0.5\textwidth]{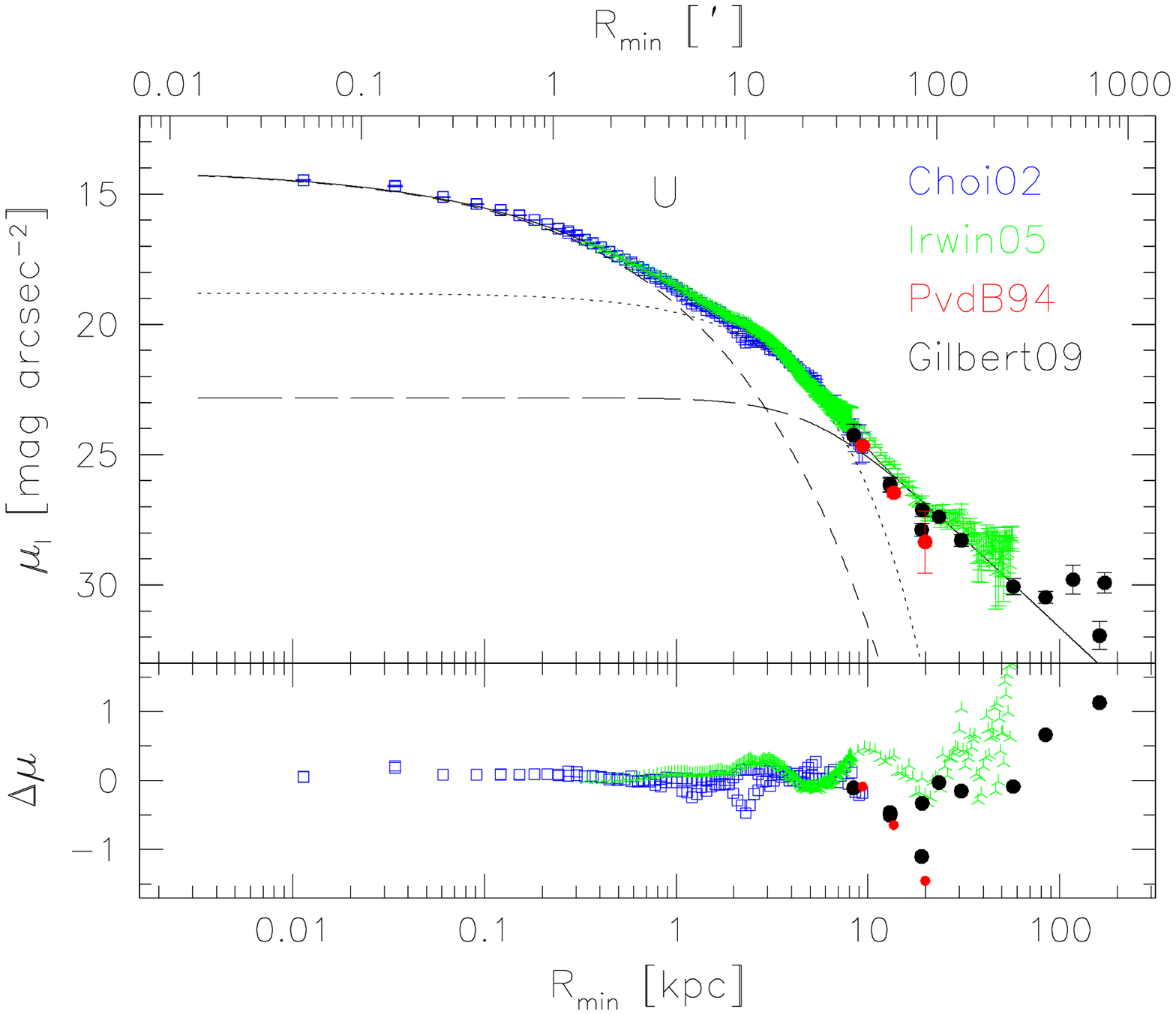} & \includegraphics[width=0.5\textwidth]{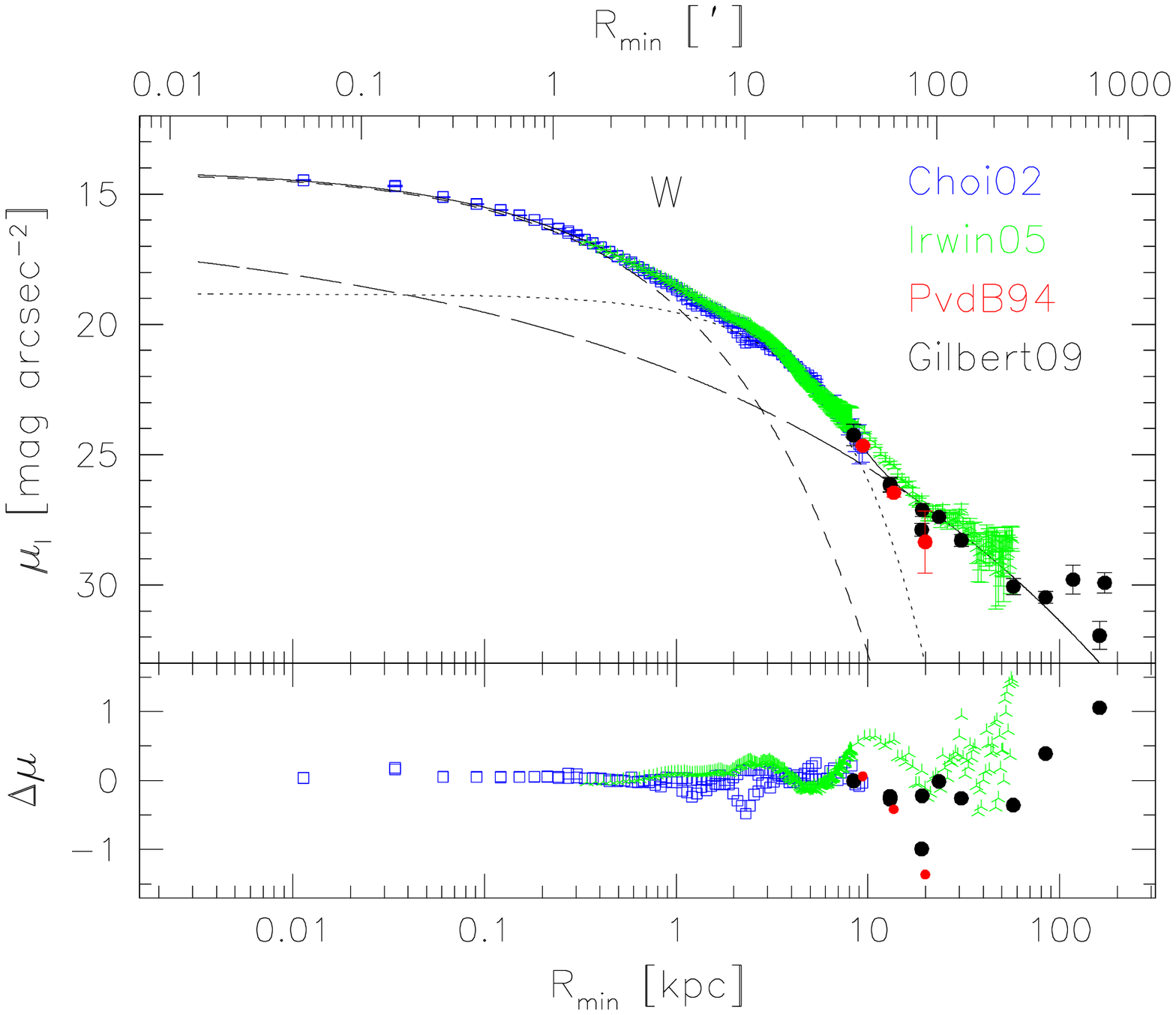}\\
\end{tabular}
\caption{Decomposition of the minor-axis composite profile for M31
with a \sersic\ bulge, exponential disk, and either a power-law halo (left)
or a \sersic\ halo (right) according to Models U and W.  The decompositions 
are consistent with the Choi \etal major axis SB profile with
$\epsilon_{\rm b} = 0.26 \pm 0.01$ and $\epsilon_{\rm d}= 0.71 \pm 0.01$.
The abscissae and parameters in Table 4 and 5 for these Models refer to the
minor axis projection.  Based on model U, the bulge is dominant within
R$_{\rm min} \ltsim 1.5$ kpc, the disk is dominant in the range $1.5 \ltsim$
R$_{\rm min} \ltsim 9$ kpc, and the halo dominates the light beyond
R$_{\rm min} \gtsim 9-15$ kpc.
}
\label{fig:M31composite_fits}
\end{center}
\end{figure*}

\newpage
\begin{figure*}
\begin{center}
\begin{tabular}{cc}
\includegraphics[width=1.0\textwidth]{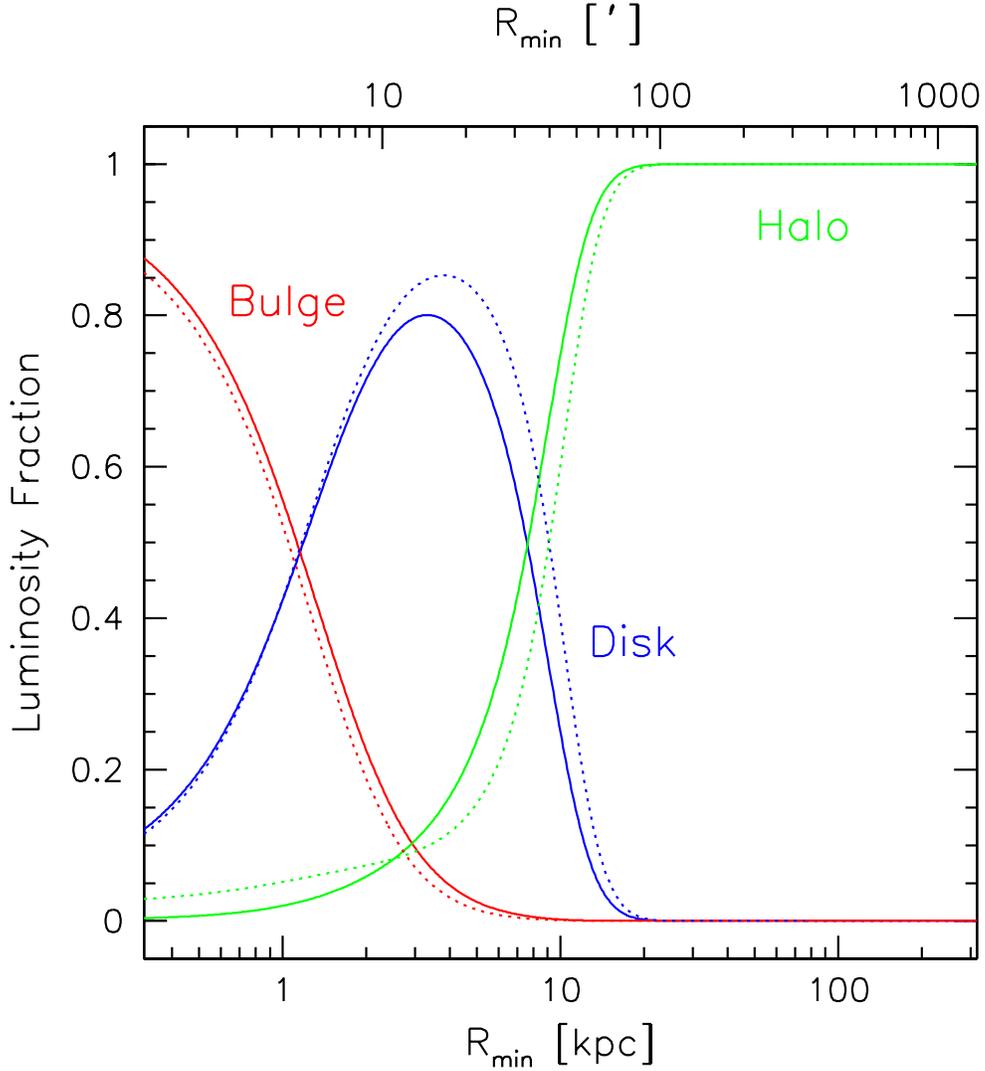} 
\end{tabular}
\caption{Total cumulative light fraction of the bulge, disk and halo components
based on the composite luminosity profile.  The solid and dotted lines represent Models
U and W, respectively.  The bulge and disk light fractions are equal just beyond
1 kpc along the minor axis; the disk and halo light fractions reach equality between
7 and 9~kpc, depending on the model.  The stellar halo is dominant beyond 9~kpc. 
}
\label{fig:lightfraction}
\end{center}
\end{figure*}

\newpage
\begin{figure*}
\begin{center} 
\begin{tabular}{cc} 
\includegraphics[width=1.0\textwidth]{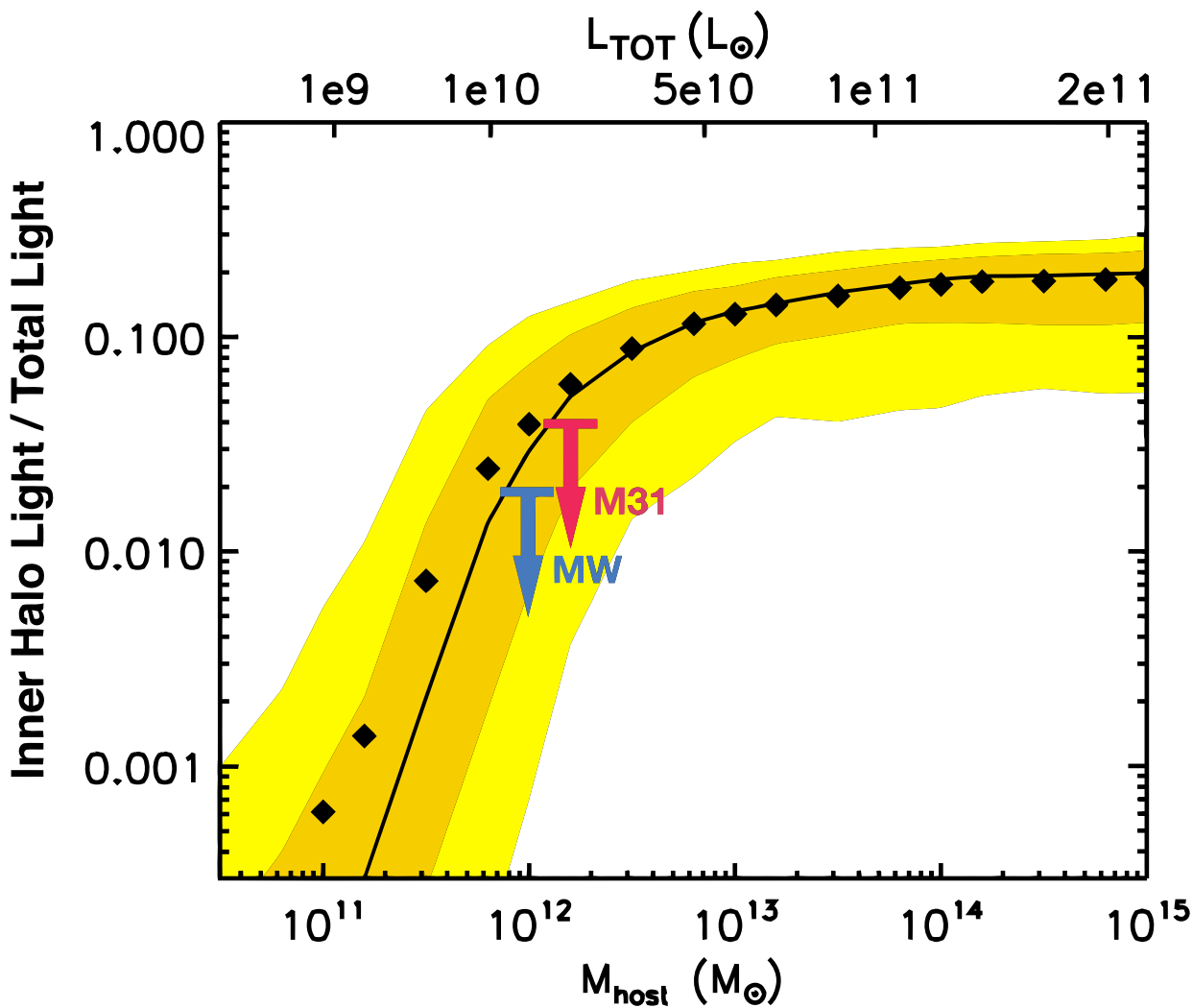} 
\end{tabular} 
\caption{Prediction of inner halo light versus total galaxy luminosity
for a range of accretion histories for host halos ranging from small
spiral galaxies ($10^{10.5}\msol$) to rich galaxy clusters ($10^{15}\msol$).
Adapted from Purcell \etal (2008; see their Figure 2).  The diamonds
denote the mean of the distribution at fixed mass based on 1000 realizations
of their analytic model, and the solid line marks the median.  The yellow
shaded regions capture the 95\% and 68\% of the distribution.  Masses for
the Milky Way (blue arrow) and M31 (red arrow) were taken from 
Courteau \& van den Bergh (1999) whereas the ratio of halo-to-total
luminosity for the Milky Way [2\%] is from Carollo \etal (2010).
The ratio of halo-to-total luminosity for M31 [4\%] is from this paper.
} 
\label{fig:HaloLight} 
\end{center} 
\end{figure*} 

\end{document}